\documentclass[onecolumn]{revtex4}    

\usepackage{graphicx}
\usepackage{mathrsfs}
\usepackage{amsmath}
\usepackage{natbib}

\begin{document}

\title{Quantifying wall turbulence via a symmetry approach. Part I. A Lie group theory}

\author{Zhen-Su She$^{1*}$, Xi Chen$^1$, Fazle
Hussain$^{1,2}$ } \affiliation{$^1$State Key Laboratory for Turbulence and Complex
Systems and Department of Mechanics,
College of Engineering, Peking University, Beijing 100871, China\\$^2$Department of Mechanical Engineering, Texas Tech University, TX,
79409-1021, USA}

\begin{abstract}
First principle based prediction of mean flow quantities of wall-bounded turbulent flows (channel, pipe, and turbulent boundary layer - TBL) is of great importance from both physics and engineering standpoints. Physically, a non-equilibrium physical principle governing spatial distribution of mean quantities is yet unknown, so that quantitative theories of technological flows are essentially empirical. Here (Part I), we present a symmetry-based approach which derives analytic expressions governing the mean velocity profile (MVP) from an innovative Lie-group analysis. The new approach begins by identifying a set of order functions (e.g. stress length, shear-induced eddy length), in analogy with the order parameter in Landau's mean-field theory, which aims at capturing symmetry aspects of the fluctuations (e.g. Reynolds stress, dissipation). The order functions are assumed to satisfy a dilation group invariance - representing the effects of the wall on fluctuations - which allows us to postulate three new kinds of invariant solutions of the mean momentum equation (MME), focusing on group invariants of the order functions (rather than those of the mean velocity as done in previous studies). The first - a power law solution - gives functional forms for the viscous sublayer, the buffer layer, the log-layer, and a newly identified central `core' (for channel and pipe, but non-existent for TBL). The second - a defect power law of form $1-r^{m}$ ($r$ being the distance from the centerline) - describes the `bulk zone' (the region of balance between production and dissipation). The third - a relation between the group invariants of the stress length function and its first derivative - describes scaling transition between adjacent layers. A combination of these three expressions yields a multi-layer formula covering the entire flow domain, identifying three important parameters: scaling exponent, layer thickness, and transition sharpness. All three kinds of invariant solutions are validated, individually and in combination, by data from direct numerical simulations (DNS).

\quad In subsequent parts, we will show the existence of a universal bulk flow constant 0.45, which asymptotes to the true Karman constant at large Reynolds numbers (\textit{Re}'s) (Part II), and an accurate description of more than 40 sets of recent experimental and numerical MVPs for channel and pipe and for \emph{Re} covering over three decades (Part III). The theory equally applies to the quantification of TBL (Part IV), and of the effects of roughness, pressure gradient, compressibility, and buoyancy, and to the study of Reynolds-averaged Navier-Stokes (RANS) models, such as $K-\omega$, with a significant improvement of prediction accuracy (Part III \& IV). These results affirm that a simple and unified theory of wall-bounded turbulence is viable with appropriate symmetry considerations.

\end{abstract}

\vspace{5pt}

\maketitle


\section{Introduction}
Canonical wall bounded flows (channel, pipe and TBL) are widely seen in both engineering applications and nature \cite{Smits2013}. Turbulent channels and pipes are internal flows driven by a pressure gradient, which fully determines the mean velocity profile and hence also friction coefficient. In contrast, TBL, driven by the freestream, develops a profile dependent on both $x$ (streamwise) and $y$ (wall-normal) coordinates. These flows are of great theoretical and practical interest and have been studied for over a century \cite{pope2000turbulent,wilcox}.

A central issue in the study of these flows is to develop viable mathematical models, in particular, to predict the mean flow properties such as the mean velocity profile (MVP), mean kinetic-energy profile (MKP), mean temperature profile (MTP), etc. Despite intensive efforts, predictions have remained essentially empirical, with the exception of the log law for MVP in the so-called overlap region. In recent decades, voluminous empirical data have been obtained from experimental and numerical studies, but have not led to any deep understanding of the principles governing the mean flow properties. Such principles, once discovered, should help to guide the statistical analysis of detailed data, offered particularly by DNS. The present work develops new theoretical concepts aiming to discover physical principles, via an innovative symmetry approach.

The study of turbulence in canonical wall-bounded flows begins by a scaling analysis focusing on the one-dimensional variation with respect to the distance from the wall \cite{pope2000turbulent}. The analysis identifies the friction velocity $u_{\tau}$, the wall viscous length unit $\delta_{\nu}\equiv\nu/u_{\tau}$, and the friction Reynolds number (\textit{Re}) $Re_{\tau}\equiv u_{\tau}\delta/\nu$ as three fundamental physical parameters, where $\delta$ is wall flow thickness (e.g. half width of the channel or radius of the pipe, or thickness of the boundary layer), and $\nu$ is the kinematic viscosity. Scaling (dimensional) analysis yields an expression for the mean velocity:
\begin{equation}\label{eq:walllaw}
U(y)=u_{\tau}\Phi(\frac{y}{\delta},\frac{y}{\delta_{\nu}})
\end{equation}
In the limit $y/\delta \to 0 $ (very close to the wall), $\Phi \left( {y/\delta ,y/{\delta _\nu }} \right) \to {\Phi _1}\left( {0,y/{\delta _\nu }} \right) = f_{w}\left( {{y^ + }} \right)$, which is called wall function, first used by \cite{prandtl1925}, and ${y^ + } = y/{\delta _\nu }$, is the distance in wall units. In the other limit $y/\delta _{\nu}\gg1 $ (very far from the wall), $\Phi \left( {y/\delta ,y/{\delta _\nu }} \right) \to {\Phi _1}\left( {y/\delta ,\infty } \right) = g\left( {y'} \right)$, which is commonly referred to as the wake function. Until now, the actual forms of $f_{w}\left( {{y^ + }} \right)$ and
$g\left( {y'} \right)$ are based on empirical propositions. The most popular model for the wall function is given by Van \cite{van1956}, which is believed to be universal for incompressible canonical wall-bounded flows, whereas the form of the wake function is more varied, depending on the geometry and other physical conditions. Specifically, a \textit{velocity-defect law} due to \citet{karman1930} reads:
\begin{equation}\label{eq:defectlaw}
U_d^ + (y/\delta ) = U_c^ +  - U_{}^ + (y/\delta ) = {F_D}(y/\delta )
\end{equation}
where  $U_d^ + $ is the mean velocity defect, $U_c^ + $ is the centerline velocity for channel and pipe flows, or velocity at the edge of TBL (typically 99\% of the freestream velocity), while the outer function ${F_D}$ is flow dependent (superscript + indicates normalization using $u_\tau$ and $\nu$).

The above two-scale description (inner and outer solutions) follows the essence of Prandtl's boundary layer concept and is commonly referred to as `classical' scaling. The celebrated log law is obtained by matching \eqref{eq:walllaw} and \eqref{eq:defectlaw}, i.e.
\begin{equation}\label{eq:loglaw}
{U^ + }({y^ + }) = \frac{1}{\kappa }\ln ({y^ + }) + B
\end{equation}
where the Karman constant $\kappa$  was believed to be universal \cite{pope2000turbulent,wilcox}, and the additive constant $B$ is flow-dependent \cite{marusic2010wall}. The log law was later derived by \citet{millikan1938} by an argument that $y^+ \partial U^+ / \partial y^+$  must be common in inner and outer solutions (matching condition). However, other quantities can be invoked to define the matching condition. For instance, if one invokes  $(y^+/U^+ )\partial U^+/\partial y^+$ as an invariant matching condition, the resulting functional form of the mean velocity being a power law. Thus, \eqref{eq:loglaw} is not the unique matching form, and that is why the debate between the log law and power law has been vivid over decades \cite{Barenblatt,Cipra17051996,Barenblatt19102004,George05}.

It has been known in the study of turbulent pipe that the log law
contradicts the boundary conditions at wall and centerline - why
Prandtl was dissatisfied with it \cite{Davidsonbook}. In 1925,
Prandtl suggested to represent effects of turbulent fluctuation
(e.g. Reynolds stress $W=-\left\langle {u'v'} \right\rangle $) in
terms of an eddy viscosity $\nu_T$ and a velocity gradient, i.e.
\begin{equation}\label{eq:stress length}
 - \left\langle {u'v'} \right\rangle  = {\nu _T}\frac{{\partial U}}{{\partial y}} = \ell _M^2{\left( {\frac{{\partial U}}{{\partial y}}} \right)^2}
\end{equation}
Here, $S=\partial U / \partial y$ is the mean shear, and ${\ell _M} =
\sqrt { W} /S $, is called the stress length function, which is essentially the mixing length introduced by \cite{prandtl1925},
presumably indicating an eddy size whose
physical interpretation, however, has been vague. Note that \eqref{eq:stress
length} is a mere definition, which requires $\ell _M$ to be
modeled. As summarized in \cite{white}, Prandtl (1925) and
von Karman (1930) took turns to make estimates of $\ell _M$, and
arrived at the following proposals:
\begin{equation}\label{eq:lmoverlap}
\text{overlap region}\quad\quad {\ell _M} \approx \kappa y
\end{equation}
\begin{equation}\label{eq:lmsublayer}
\text{sublayer}\quad\quad {\ell _M} \approx y^2
\end{equation}
\begin{equation}\label{eq:lmouter}
\text{outer layer}\quad \quad {\ell _M} \approx const.
\end{equation}
The linear assumption \eqref{eq:lmoverlap} leads to the log law, while \eqref{eq:lmsublayer} is proposed to satisfy wall condition, i.e., $\ell_M\rightarrow0$ as $y\rightarrow0$. Various combination of \eqref{eq:lmoverlap}-\eqref{eq:lmouter} yields formulas for wall function and wake function. For example, by assuming both \eqref{eq:lmoverlap} and \eqref{eq:lmsublayer}, van Driest (1956) proposed an exponential damping function:
\begin{equation}\label{eq:lmdamp}
{\ell _M} \approx \kappa y[1 - \exp ( - {y^ + }/A)]
\end{equation}
where $A \approx 26$ is determined for a flat plate TBL. One may also combine \eqref{eq:lmdamp} with \eqref{eq:lmouter} with a matching condition, to produce a desired functional form covering both inner and outer MVP - widely used in RANS models \citep{pope2000turbulent,wilcox}. Another well-known model was suggested by \citet{coles1956} going from the overlap region (the log law) to the outer region, by taking into account both \eqref{eq:defectlaw} and \eqref{eq:loglaw},
\begin{equation}\label{eq:coles}
{U^ + }({y^ + }) = \frac{1}{\kappa }\ln ({y^ + }) + B + \frac{{2\Pi }}{\kappa }f(\frac{y}{\delta })
\end{equation}
with $\Pi$, called Cole's wake parameter, and $f(x)$, the wake function. A model widely used for pipe or channel or TBL is: $f(x)= sin^{2}(\pi x/2)$. It is easy to show that \eqref{eq:coles} yields a more complicated function than \eqref{eq:lmouter}, but corresponds still to a model similar to \eqref{eq:lmoverlap}-\eqref{eq:lmouter} in spirit.

However, as we show below, the correct scaling in the viscous sublayer layer is ${\ell _M} \propto y^{3/2}$, and \eqref{eq:lmsublayer} describes the scaling in a buffer layer. In addition, \eqref{eq:lmouter} is also incorrect for pipe and channel near the centerline, due to the presence of a core layer. Thus, the classical mixing length theory is not only empirical in nature, but also sometimes incorrect. That it is widely accepted is because its errors are tolerable in most applications, and DNS data for its verification become available only recently. The crucial defect is the missing derivation: since the physical principle behind the suggested \emph{law of wall} ($f_{w}$) and \emph{velocity-defect law} ($F_{D}$) is unknown, it is not possible to define the domain of validity and to extend/modify the functions to include other physical effects such as pressure gradient and roughness. Two important issues remain: how exact is the log law, and how universal is the Karman constant $\kappa$? The log law has been questioned by \citet{Barenblatt}, \citet{Barenblatt19102004} and \citet{George05}; they argue that the power law is more natural and fits the MVP data in a wider domain. Also, $\kappa$ has been assumed to be a universal constant for a long time \cite{pope2000turbulent,wilcox}, equaling 0.40$\sim$0.41. However, as more data accumulate, $\kappa$ measured with the classical definition of the log law \eqref{eq:loglaw} show a 20\% variation - from 0.37 to 0.45 - which, with no sign of convergence, is rather frustrating \cite{marusic2010wall,Smits,Alfredsson2013}. The present work intends to resolve these controversies.

Several recent studies are noteworthy. \citet{monkewitz2007}, \citet{nagib2008} (hereafter cited as MCN), after a tremendous effort to parameterize entire MVP of channel, pipe and TBL, claim that the classical description (an inner-outer two layer model with a logarithmic overlap region) is better than the competing power law model. However, more than ten fitting parameters, defying any physical explanation, are needed for each canonical flow. Moreover, the choice of wake functions has no sound physical basis, and $\kappa$ as a free fitting parameter is called `Karman coefficient'. Another model, by \citet{nickels2004}, employs a three-layer description (with an explicit logarithmic layer), but the switch from channel and pipe flows to TBL has no justification. In addition, in Nickels's model, $\kappa$ is also a fitting parameter with no discussion of its universality, and with no attempt to determine wake parameters to predict the full MVP. \cite{javier2006} proposed also a model for turbulent eddy viscosity, following \citet{Reynolds1972}, which leads to a closure description of the MVP in channel flows, but no intent to generalize to TBL is reported. A physical model by \citet{lvovprl} addresses explicitly effects of turbulent fluctuations on the mean velocity using three characteristic length scales, but then still employs an empirical wake function (by inspecting DNS data) for describing channel and pipe flows, and no attempt to extend the analysis to TBL is reported. In summary, all quantitative theories \cite{nickels2004, javier2006, monkewitz2007, Panton2007, lvovprl} for wall turbulence remain essentially empirical.

The universality or otherwise of the Karman constant, and the scaling of mean flow, are listed as outstanding questions in the recent review article of \citet{marusic2010wall}. \citet{Smits} suggest to develop new facilities with improved measurement accuracy, and in particular, to open new theoretical perspectives. A proposed specific direction is to develop a complete description of MVP like MCN model, but with more physical contents and better theoretical underpinning and hopefully comparing with more data. Our work pursues this line of approach.

The question posed here is: does there exist a physical principle to determine the form of $f_{w}(y^{+})$ and $g(y'$), or the entire function $\Phi(y',y^{+})$? The present symmetry-based analysis offers a positive answer, at least for canonical wall turbulence, with evidence as below. In this paper (Part I), we introduce the new theoretical framework of a symmetry analysis using a set of special quantities, called order function, which is an extension of order parameter in Landau's mean-field theory (Landau, 1958). It is used to represent the symmetry property of the flows, see Appendix A for details. A specific candidate for the order function is the stress length function which is assumed to possess an invariance when the RANS equation is under dilation group of transformation. We then carry out a Lie-group analysis on the (unclosed) mean momentum equation (MME) and construct a series of new invariant solutions. However, it is different from previously reported Lie-group analysis of the RANS equation by \cite{Oberlack2001}, \cite{lindgren2004} and \cite{marati2006mean}, in which the mean velocity is systematically taken as the invariant variable. In our view, due to fluctuations, the mean velocity itself has broken most Lie group invariance, which is at best restored in very restricted domains. This is the origin of the difficulty to construct global invariant solutions \cite{Oberlack2010new}. However, by changing the symmetry variable - from the mean velocity to the stress length (order) function - dilation symmetries for the latter are identified in the present work, which lead to a complete description of the entire MVP.

In a nut shell, current derivation of the MVP involves three steps. First, the stress length function is assumed to be the key symmetry variable, representing the effects of fluctuations (e.g. Reynolds stress) on the mean flow. Second, a dilation group analysis is applied to the MME, yielding group invariants for the stress length function (and its derivative), which further leads to the construction of dilation invariant solutions (three kinds). Third, a multi-layer structure is assumed; each layer corresponds to one balance mechanism in the turbulent kinetic-energy equation. The transition from one layer to another is assumed to satisfy a generalized Lie-group invariance, satisfying the continuity condition so that the matching technique yields a complete analytic formula for the stress length function over the entire flow domain, hence the entire MVP. This is the main content of this paper.

In subsequent parts of this series, we will show the existence of a universal bulk flow constant 0.45, which asymptotes to the true Karman constant at large \textit{Re}'s (Part II), and an accurate description of more than 40 sets of recent experimental and numerical MVPs for channel and pipe and for \emph{Re} covering over three decades (Part III). The theory equally applies to the quantification of TBL (Part IV), and of the effects of roughness \cite{shenjp}, pressure gradient, compressibility, and buoyancy, and to the study of RANS models, such as $K-\omega$, with significant improvement of prediction accuracy (Part III \& IV). These results affirm that a simple and unified theory of wall-bounded turbulence is viable with appropriate symmetry considerations.

An important result is a sound assessment of the vivid debate between the log law and power law. Both log law and power law are local descriptions of MVP in a restricted domain, although the power law description improves the log law in a marginally larger domain. We show (Part III) that the log-law is asymptotically correct, not only in the MVP description, but also for the description of the centerline velocity, the bulk velocity (the volume averaged flux), and the friction coefficient. At moderate \textit{Re}'s, the overlap region of the log law is very narrow, but our multi-layer description yields the correct description over the entire flow domain, which asymptotes to the log law, so that the integrated quantities (e.g. the bulk velocity, friction coefficient) follow definitively the log law scaling. The validity of the log law is further supported by the universal Karman constant. Thus, it is reasonable to call for a close of this debate, as both channel and pipe data at moderate and large \textit{Re}'s are all accurately described under a unified setting. The extension to TBL yields similar results, as reported in Part IV.

The current approach differs from the classical mixing length theory in several ways. First, the current stress length is taken under a general concept of the order function, whose functional form is motivated from (dilation) symmetry property of the flow, but not modeled by any ad hoc formula. Thus, its analytic expression is not arbitrary. Second, the multi-layer structure is physically sound; although not new in its concept, its analytic definition through the scaling transition of $\ell _M$ has led to the quantification of all layers and has helped to identify the effects of channel and pipe geometries with different integer exponents, as well as the differences between the internal and external flows (e.g. TBL, see Part IV). This connection to the actual physical (multi-layer) structure is missed in the classical mixing length theory. Third, and most importantly, the current symmetry principle allows us to generalize the analysis to other situations including roughness, compressibility, pressure gradient, and buoyancy, etc., unlike previous empirical models (i.e. van Driest¡¯s damping function, or the MCN formula), since the dilation invariance is likely the symmetry principle for all wall-bounded flows. Finally, note that the stress length function is just one of the characteristic length functions (order functions) to display symmetry in turbulence. When more complex physical conditions occur (e.g. with density and temperature fluctuations), new order functions will appear, but current theoretical framework will prevail. Thus, the current theory is significantly advanced compared to the classical mixing length theory.

This paper (Part I) is organized as follows. In section II, we summarize previous studies using Lie group symmetry analysis and introduce our study of invariant solution of stress length function. Section III presents three kinds of dilation invariant solutions of stress length function. In section IV a composite solution for the entire flow domain is derived (using the multiplicative rule), and shown to agree well with DNS data. In section V, we summarize and further discuss the results. In Appendix A, we present the main features of the new symmetry based approach. In Appendix B, we introduce the reader to the essential concepts in Lie group symmetry analysis, and no previous knowledge on Lie group is assumed; for more exhaustive discussions, see \citet{bluman1991,Cantwellbook}.

\section{Symmetry approach to the study of wall flows}\label{sec:2}

Symmetry is one of the most important concepts in physics
\cite{kadanoff2009more, falkovich2009}. In mathematics, it is
defined as a patterned self-similarity that can be demonstrated with
the rules of operation. Generally, symmetry implies invariance. If
there is an invariant in the system which keeps it unchanged under
transformations, we say that there exists a symmetry. Lie groups are
basic tools to characterize continuous symmetry in mathematical
structures as well as in nature. It is originally developed by
Sophus Lie at around 1890s \cite{bluman1991,Cantwellbook}, laying
the foundation of the theory of continuous transformation groups,
and now provides a framework for analyzing the continuous symmetries
of differential equations. In general, Lie group symmetry analysis
is used to reduce the differential order or the number of
independent variables; sometimes, it yields a way to construct an
explicit solution for an ordinary differential equation (ODE) or a
partial differential equation (PDE).

Applying Lie group symmetry to study fluid dynamics is under-developed. Early studies devoted to Lie group symmetry analysis for the Navier-Stokes (NS) equations, i.e.
\begin{equation}\label{eq:nsdensity}
\frac{\partial u_{k}}{\partial x_{k}} =0
\end{equation}
\begin{equation}\label{eq:nsmom}
\frac{{\partial {u_i}}}{{\partial t}} + {u_k}\frac{{\partial {u_i}}}{{\partial {x_k}}} = \nu \frac{{{\partial ^2}{u_i}}}{{\partial x_k^2}} - \frac{{\partial p}}{{\partial {x_i}}}
\end{equation}
(note that the density is absorbed in $p$) and the relevant symmetry transformations can be found in \citet{Frischbook} and \citet{Cantwellbook}, which are briefly summarized as below:
\begin{equation}\label{eq:symtrans}
\begin{split}
& (a)   \text{Space-translations: }   t^{*}=t,x_{i}^{\ast}=x_{i}+\epsilon_{i},u_{i}^{\ast}=u_{i},p^{\ast}=p \\
& (b)   \text{Time-translations: }   t^{*}=t+ \epsilon,x_{i}^{\ast}=x_{i},u_{i}^{\ast}=u_{i},p^{\ast}=p \\
& (c)   \text{Galilean transformation: }   t^{*}=t,x_{i}^{\ast}=x_{i}+\epsilon_{i}t,u_{i}^{\ast}=u_{i}+\epsilon_{i},p^{\ast}=p \\
& (d)   \text{Rotations: }   t^{\ast}=t,x_{i}^{\ast}=a_{ij}x_{j},u_{i}^{\ast}=a_{ij}u_{j},p^{\ast}=p \\
& (e)   \text{Dilations: }   t^{\ast}=e^{\epsilon}t,x_{i}^{\ast}=e^{\lambda \epsilon}x_{i},u_{i}^{\ast}=e^{(\lambda -1)\epsilon}u_{i},\nu^{\ast}=e^{(2\lambda -1)\epsilon}\nu,p^{\ast}=e^{(2\lambda -1)\epsilon}p
\end{split}
\end{equation}
Here $\epsilon$ (and $\epsilon_{i}$)$\in R $ denotes the Lie group parameter; $a_{ij}$ is the element of a unit orthogonal matrix (the reflection symmetry is also included); $\lambda \in R$ is a free parameter for dilations -(dilation on viscosity also; $\lambda =1/2$ if no dilation of viscosity).

Symmetry analysis can indeed provide insights on the solution of the NS equation. The benefit of introducing symmetry transformation groups is: if ${u_i}(t,{x_i})$ is a solution for a velocity field, then
$u_i^*({t^*},x_i^*)$ also satisfies NS equation, and may be a solution provided that the boundary conditions also remain invariant under the transformation (note that velocity field is determined by the governing equation and its boundary conditions). Then, the velocity field ${u_i}(t,{x_i})$ is said to have the following symmetries for each of the above items in \eqref{eq:symtrans}: homogeneity in space and time (a-b); independent of reference frame, both velocity and acceleration permitted (c); isotropy: isotropic turbulence with zero mean velocity (d); and \textit{Re} similarity in (e) for $\lambda =1/2$ (since no changing on viscosity).

However, boundary conditions may break the above symmetries; for
example, wall proximity breaks the isotropy (d); the translation
symmetry (a) is broken in the normal direction as any change in
normal location changes $u$. It may also introduce new symmetries
under specific conditions (such as periodic condition or plane
symmetry condition). In 1970s, a turbulent spot was studied
through a symmetry analysis for the two-dimensional, unsteady,
stream-function equation, where the stream functions are expressed in similarity variables
(group invariants), and the unsteady particle displacements are
reduced to an autonomous system in the plane of similarity
variables \cite{cantwell1978a,cantwell1978b}. These two papers are the first examples using symmetry analysis for real wall flows.

Before presenting our analysis, we introduce a ground material \cite{Cantwellbook}: using Lie group symmetry analysis to re-derive the invariant solutions for laminar boundary layer (i.e. Blasius equation). This offers a `classical' way to reduce PDEs to ODEs, and to obtain the invariant solution. It is also called as the similarity solution, and more examples can be found in \cite{Cantwellbook}.

For the laminar boundary layer (2D), the streamwise viscous diffusion in the momentum equation is small and ignored (exactly zero for a parallel flow). Thus the mass and momentum equations are approximated to
\begin{equation}\label{eq:lamdensity}
\frac{{\partial u}}{{\partial x}} + \frac{{\partial v}}{{\partial y}} = 0
\end{equation}
\begin{equation}\label{eq:lammom}
\frac{{\partial u}}{{\partial t}} + u\frac{{\partial u}}{{\partial x}} + v\frac{{\partial u}}{{\partial y}} = \nu \frac{{{\partial ^2}u}}{{\partial {y^2}}} - \frac{{\partial p}}{{\partial x}}
\end{equation}
which is the laminar boundary layer (BL) equation. Introducing the streamfunction $\psi$: $u = {{\partial \psi }}/{{\partial y}}$ and $v =  - {{\partial \psi }}/{{\partial x}}$, \eqref{eq:lammom} is rewritten as
\begin{equation}\label{eq:lammomstream}
\frac{{\partial \psi }}{{\partial y}}\frac{{{\partial ^2}\psi }}{{\partial x\partial y}} - \frac{{\partial \psi }}{{\partial x}}\frac{{{\partial ^2}\psi }}{{\partial {y^2}}} - \nu \frac{{{\partial ^3}\psi }}{{\partial {y^3}}} = 0
\end{equation}
where time derivative and pressure gradient terms are eliminated under steady flow and zero-pressure-gradient (ZPG) conditions. Accordingly, we can find dilation symmetry transformation for \eqref{eq:lammomstream}, with arbitrary dilations on space coordinate and $\psi$:
\begin{equation}\label{eq:symtranstreamf}
x^{\ast} =e^{\lambda_{1}\epsilon}x,\quad y^{\ast} =e^{\lambda_{2}\epsilon}y,\quad \psi^{\ast} =e^{\lambda_{3}\epsilon}\psi
\end{equation}
Substituting \eqref{eq:symtranstreamf} into \eqref{eq:lammomstream}, the invariance condition yields the relation
\begin{equation}\label{eq:2.8}
\lambda_{3}=\lambda_{1}-\lambda_{2}
\end{equation}
Normalized with parameter $\lambda_{1}$ (i.e. $\lambda=\lambda_{2}/\lambda_{1}$, $\lambda_{3}/\lambda_{1}=1-\lambda$), the dilation symmetry transformation is:
\begin{equation}\label{eq:dilationtransformation}
x^{\ast}=e^{\epsilon}x,\quad y^{\ast}=e^{\lambda \epsilon}y,\quad \psi^{\ast}=e^{(1-\lambda)\epsilon}\psi
\end{equation}
Note that\eqref{eq:dilationtransformation} indicates different dilations on $x$ and $y$, different from the homogeneous dilations for three space coordinates, i.e. (e) in (2.3). The reason is that we neglect the small streamwise viscous diffusion term in the laminar boundary layer (also applies for TBL). If we keep the viscous diffusion term in $x$, then $\lambda=1$, i.e. the same dilations in $x$ and $y$ .

Furthermore, the boundary condition also needs invariant under dilation, that is
\begin{equation}\label{eq:boundarydilation}
\frac{{\partial \psi }}{{\partial y}}{|_{y = \infty }} = {U_\infty } = \frac{{\partial {\psi ^ * }}}{{\partial {y^ * }}}{|_{{y^ * } = \infty }}
\end{equation}
Then, substituting \eqref{eq:dilationtransformation} into \eqref{eq:boundarydilation} yields $\lambda=1/2$, and the symmetry transformation is
\begin{equation}\label{eq:dilationtrans2}
x^{\ast}=e^{\epsilon}x,\quad y^{\ast}=e^{ \epsilon /2}y,\quad \psi^{\ast}=e^{\epsilon/2}\psi
\end{equation}
Two dilation invariants of \eqref{eq:dilationtrans2} are thus (combing $y$ and $x$, $\psi$ and $x$ by eliminating $\epsilon$)
\begin{equation}\label{eq:liegroupinv}
\pmb{\rm{I}}_{1} = {y}/{{\sqrt x }};\quad \quad \pmb{\rm{I}}_{2} = {\psi }/{{\sqrt x }}
\end{equation}
Note that $\pmb{\rm{I}}_{1}$ and  $\pmb{\rm{I}}_2$, being independent of $\lambda$, are dimensional, and combination of $\pmb{\rm{I}}_{1}$ and $\pmb{\rm{I}}_{2}$ is also an invariant.

Invariant solution of \eqref{eq:lammomstream} is denoted in a general form $\Omega (\pmb{\rm{I}}_{1},\pmb{\rm{I}}{_2}) = 0$. In reality, substituting \eqref{eq:liegroupinv} into the original PDE \eqref{eq:lammomstream} yields an ODE
\begin{equation}\label{eq:liegroupinveq}
\frac{{d^3}\pmb{\rm{I}}_{2}}{{d{\pmb{\rm{I}}_1}^3}} + \frac{{{\pmb{\rm{I}}_2}}}{{2\nu }}\frac{{{d^2}{\pmb{\rm{I}}_{2}}}}{{d{\pmb{\rm{I}}_{1}}^2}} = 0
\end{equation}
which is a specific expression for $\Omega ({{\pmb{\rm{I}}}_1},{{\pmb{\rm{I}}}_2}) = 0$, and can be solved numerically. Note that it also can be obtained by a dimensional analysis. In fact, dimensional analysis postulates the set of similarity variables through trial and error. In contrast, symmetry analysis applied to ODEs or PDEs defines similarity variables by identifying group invariants and is therefore more systematic. Both approaches need a physical insight of the problem, and can be used together. For example, one can further define dimensionless variables $\eta$ and $f$ by normalizing $\pmb{\rm{I}}_{1}$ and $\pmb{\rm{I}}_{2}$, inspired by above group invariants, that are
\begin{equation}\label{eq:liegroupdimensionlessinv}
\eta  = {y}/{{\sqrt {x\nu /{U_\infty }} }};\quad \quad f = {\psi }/{{\sqrt {x\nu {U_\infty }} }}
\end{equation}
then \eqref{eq:liegroupinveq} is transformed into the well known Blasius equation:
\begin{equation}\label{eq:blasiuseq}
\frac{{{d^3}f}}{{d{\eta ^3}}} + \frac{f}{2}\frac{{{d^2}f}}{{d{\eta ^2}}} = 0
\end{equation}
The above case shows how we use physical insights to obtain a new symmetry, i.e. \eqref{eq:dilationtrans2}, compared to those \eqref{eq:symtrans} for the NS equations, and how symmetry analysis leads to invariant (similarity) solutions. For more laminar flow examples, see \cite{Cantwellbook}.

Now, let us explain the meaning of dilation symmetry in the laminar boundary layer. Consider a solution $u$ at $(x,y)$ for the BL equation. According to the symmetry transformation, $(x,y,u)$ are transformed to $(x^{\ast},y^{\ast},u^{\ast})$. Note that $u^{\ast}$, obtained purely by the transformation, satisfying the same BL equation, is expected but not necessarily a solution at the position $(x^{\ast},y^{\ast})$, because it has to satisfy boundary condition also. So assume that the correct solution at the position $(x^{\ast},y^{\ast})$ is $u^{S}$ (in this case is the numerical Blasius solution). Then, there is a symmetry for the velocity field $u$ only if $u^{\ast}=u^S$; otherwise symmetry is broken. Accordingly, the dilation defined in \eqref{eq:dilationtrans2} is a symmetry transformation, and the result of this dilation happens to also satisfy $u^{S}=u^{\ast}$; hence there is a dilation symmetry in the laminar boundary layer.

For TBL, we need to treat the unclosed RANS equations, which read:
\begin{equation}\label{eq:ransdensity}
\frac{{\partial {{\bar u}_k}}}{{\partial {x_k}}} = 0
\end{equation}
\begin{equation}\label{eq:ransmom}
\frac{\partial \bar{u}_1}{\partial t} + \bar {u}_k\frac{\partial \bar {u}_1}{\partial {x_k}} + \frac{\partial \overline{u'_1 u'_k}}{\partial {x_k}} = \nu \frac{\partial ^2\bar {u}_1}{\partial {x_k}\partial {x_k}} - \frac{\partial \bar p}{\partial {x_1}}
\end{equation}
(where $\bar{u}_1$ is the streamwise mean velocity). It is easy to verify that all of the NS symmetries listed in \eqref{eq:symtrans} are admitted by \eqref{eq:ransdensity} and \eqref{eq:ransmom}, after Reynolds decomposition. \cite{Oberlack2001} has carried out the Lie group analysis of the RANS equations and defined the invariant solution by using the invariant of the mean velocity, i.e.
\begin{equation}\label{eq:invOberlack}
G(\bar{u}_1,y)=0   \quad\Rightarrow\quad  G(\bar{u}_1^{\ast},y^{\ast})=0
\end{equation}
which is equivalent to,
\begin{equation}
G(\pmb {\rm I})=0
\end{equation}
where $\pmb{\rm I} $ is a group invariant and assumed to be constant.
A specific candidate invariant solution is the log law, which is obtained by postulating the dilation with parameter $\lambda=1$, combined with a translation in $\bar{u}$ (i.e. the streamwise mean velocity $\bar{u}_1$):
\begin{equation}\label{eq:2.29}
x^{\ast}=e^{\epsilon}x, \quad y^{\ast}=e^{\epsilon}y,\quad \bar{u}^{\ast}=\bar{u}+\epsilon/\kappa
\end{equation}
In this case, the group invariant (by eliminating $\epsilon$) is
\begin{equation}\label{eq:2.30}
{\pmb{\rm{I}}} = {\bar{u}^*} - \ln {y^*}/\kappa  = \bar{u} - \ln y/\kappa
\end{equation}
Then, by assuming $\pmb{\rm I}=const.$ one obtains the log law:
\begin{equation}\label{eq:2.31}
\bar{u} = \frac{1}{\kappa }\ln y + {\pmb{\rm{I}}}
\end{equation}
However, the translation in $\bar{u}$ is questionable, since it breaks the no-slip wall condition, i.e. $\bar{u}=0$ at $y=0$. It is important to note that a symmetry based approach to describe mean quantities must use a transformation group which is rigorously (or at least asymptotically) valid, since symmetry reveals universal principle governing the MVP. The dilation group is the only rigorous invariance group of wall turbulence in the presence of wall. All our subsequent results are thus based on the dilation group.

Note that the RANS equations are unclosed. The unclosedness stems from the presence of fluctuations, which introduces a major challenge to the symmetry analysis due to arbitrariness in specifying invariant solution manifold. We undertake this challenge by closely relying on numerical solutions of the NS equations (e.g. DNS data) to verify if a theoretical proposal is valid or not. The invariant solutions derived in the subsequent analysis should be understood as a set of functions, which lie on the invariant manifold leaving the RANS equations unchanged under a set of Lie group transformations and which cannot be distinguished from true empirical solutions. This concept, although not mathematically rigorous, is physically transparent: each proposed invariant solution coincides with an empirical statistical solution of the original NS equations (i.e. MVP). It turns out that correct invariant solutions are obtained by assuming the dilation invariance of the stress length function.

\section{Symmetry analysis with length order functions}
{Current idea about the dilation symmetry in the wall turbulence can be summarized as follows.} Only one kind of symmetry - dilation symmetry - is permitted in the presence of a no-slip wall. Symmetries in the RANS equations are broken in velocity, but preserved in length order functions, {as the latter display the effects of the wall on the symmetry properties of fluctuations. It is indeed verified by DNS data} that with increasing $y$, length order functions display different scaling, i.e. different exponents for dilations. Each spatial domain obeying this scaling property in the length function is referred to as a layer, where a well-defined energy balance mechanism is found. And the combination of these layers is the multi-layer structure of wall-bounded flows. For transit from one layer to another, a generalized dilation invariant involving the dilation {invariants} of the length order function and of its gradient is postulated and shown to describe well the transitions between the layers.

Let us take a canonical channel flow in the $x$ direction, for example. The mean momentum equation has the following form, steady in time,
\begin{equation}\label{eq:3.2}
 \frac{\partial }{{\partial y}}{\left( {\ell _M^{}\frac{{\partial \bar u}}{{\partial y}}} \right)^2} + \nu \frac{{{\partial ^2}\bar u}}{{\partial {y^2}}} + \bar{P}=0
\end{equation}
where $\bar {P}$ is the constant pressure gradient driving the channel flow, and the nonlinear mean convection term, the diffusion terms in $x$ and $z$ directions are all zero. In \eqref{eq:3.2}, the Reynolds stress is replaced by the stress length function. It is easy to show that both the mean continuity equation (trivially zero) and the mean momentum equation \eqref{eq:3.2} have a following dilation symmetry transformation:
\begin{equation}\label{eq:3.3}
\begin{array}{l}
\quad y^* = e^{\varepsilon}y,\quad \ell_M^* = e^{\alpha\varepsilon }\ell _M,\quad \bar {u}^* = e^{\beta \varepsilon}\bar {u}\\ \nu ^* = e^{(\beta  + 2\alpha  - 1)\varepsilon}\nu,\quad \bar {P}^* = e^{(2\alpha  + 2\beta  - 3)\varepsilon}\bar P
\end{array}
\end{equation}
Note that in \eqref{eq:3.3} we introduce a separate dilation scaling exponent $\alpha$ for the length function; this is distinct from that derived from dimensional analysis, which gives $\alpha\equiv1$. This last result follows from the definition of the Reynolds stress:
\begin{equation}\label{eq:3.6}
{\overline {u'v'} ^*} =  - {[\ell _M^*{(\partial \bar u^*/\partial y^*)}]^2} = {e^{2(\alpha  + \beta  - 1)\epsilon}}\overline {u'v'},
\end{equation}
which, if taking the normal scaling argument, should equal to $e^{2\beta\epsilon}\overline {u'v'}$, i.e. $\alpha =1$. However, according to our understanding, the presence of fluctuations has broken the dilation invariance of the mean velocity (i.e. $\beta$ is no longer constant). Instead, a random dilation of velocity should be considered (see below), which yields $\alpha\neq 1$. Then, it may be possible that the dilation of $\ell_M$ is locally preserved (our assumption) in each layer of the multiple layers (sublayer, buffer layer, log layer, core layer, etc.). In other words, introducing $\alpha$ opens the possibility to characterize different layers via the present symmetry approach.

This marks a notable departure from \cite{Oberlack2001}, which is of major significance for symmetry analysis of turbulent flows, as the dilations for the mean and fluctuations are now treated separately. A similar account is made by Kraichnan regarding the random Galilean transformation for the NS equation \cite{Frischbook}, i.e. $x_i ^\ast =x_i +b_i t$, $u_i ^\ast =u_i+b_i$, where $b_i$ is random and isotropically (say, Gaussian) distributed. In this case, there is no translation for the mean velocity $\bar{u}_i$, since $\bar{b}_i =0 $; but there is a translation on the fluctuation, i.e. $u'^{\ast}=u'+b_i$; in other words, fluctuation and mean are transformed differently. Similarly, if we introduce a random dilation transformation for the NS equation, i.e. $u_i^{\ast}=\lambda u_i$, where $\lambda$ is random, we obtain, ${\bar{u}_i}^{\ast}=\overline{\lambda u_i}$, and $u'^{\ast}_{i}=u^{\ast}-{\bar{u}_i}^{\ast}=\lambda u_i -\overline{\lambda u_i}$. Taking the parallel flow for example, let $\lambda$ be independent of $u_i$, then $\bar{u}^{\ast}=\bar{\lambda} \bar{u}$, $u'^{\ast}=\lambda u-\bar{\lambda }\bar{u}$, and the Reynolds stress is transformed as $\overline{u'v'}^{\ast}=\overline{\lambda^2} \overline{u'v'}$. Since $\overline{\lambda^2}$ is generally different from $\overline{\lambda}^2 $, the mean and fluctuations are also transformed differently. This possibility was not considered in prior Lie-group analysis of the RANS equation \cite{Oberlack2001,lindgren2004,marati2006mean}.

Below, we will use the dilation invariance of $\ell_M$ (and its gradient) under \eqref{eq:3.3}, i.e.
\begin{equation}\label{eq:3.5}
F(\ell_{M},\partial \ell_{M}/\partial y,y)=0 \quad \Rightarrow \quad F(\ell_{M}^{\ast},\partial \ell_{M}^{\ast}/\partial y^{\ast},y^{\ast})=0
\end{equation}
to construct invariant solution of \eqref{eq:3.2}.
The specific expression for $F$ will be given later (by postulating the constancy of dilation invariants). Once the length function is known, the mean velocity can be solved from \eqref{eq:3.2}.
In other words, the dilation invariant of $\ell_M$ offers an
additional constraint to yield a solution for the unclosed problem of turbulence.

Before presenting specific invariant solutions, we extend above analysis to TBL. One way is to assume parallel flow as done by \citet{Oberlack2001}, which yields the same dilation symmetry for TBL as for pipe and channel. Here, we present an alternative way, namely neglecting the streamwise viscous diffusion, like that in laminar boundary layer. Then, the steady streamwise mean momentum equation reads
 \begin{equation}\label{eq:3.7}
\bar u\frac{\partial \bar u}{\partial x} + \bar v\frac{\partial \bar u}{\partial y} + \frac{\partial \overline {u'v'} }{\partial y} = \nu \frac{\partial ^2\bar u}{\partial y^2} +\bar P
 \end{equation}
and we introduce the stress length function, to obtain:
\begin{equation}\label{eq:3.8}
\bar u\frac{{\partial \bar u}}{{\partial x}} + \bar v\frac{{\partial \bar u}}{{\partial y}} = \frac{\partial }{{\partial y}}{\left( {\ell _M^{}\frac{{\partial \bar u}}{{\partial y}}} \right)^2} + \nu \frac{{{\partial ^2}\bar u}}{{\partial {y^2}}} +\bar P
\end{equation}
It is easy to verify that the dilations permitted by both \eqref{eq:ransdensity} and \eqref{eq:3.8} are
 \begin{equation}\label{eq:3.9}
\begin{array}{l}
x^* = e^{(3-2\alpha)\epsilon}x,\quad y^*=e^{\epsilon}y,\quad \ell _M^* = e^{\alpha \epsilon}\ell _M,\quad \bar{u}^*=e^{\beta \epsilon }\bar{u}\\ \bar{v}^*=e^{(\beta+2\alpha  - 2) \epsilon }\bar{v},\quad \nu ^*= e^{(\beta  + 2\alpha  - 1)\epsilon}\nu,\quad \bar P = e^{(2\alpha+2\beta-3)\epsilon}\bar P
\end{array}
 \end{equation}
Note that \eqref{eq:3.9} shows different dilations on $x$ and $y$, similar to that in the laminar boundary flow.

Now we postulate three kinds of invariant solutions specific to the dilation symmetry of the stress length function. The stress length is regarded as the order function for the following reason. Inspecting DNS data indicates that the stress length increases from zero at the wall, and undergo changes through different layers with different scaling exponents, which is exactly the property of the order parameter in the mean field theory: displaying symmetry variation during a phase transition \cite{kadanoff2009more}. Thus, $\ell$ (indicates $\ell_M$) is a proper candidate for order function to capture different layers. Noting that \eqref{eq:3.8} involves the first derivative of $\ell$, we use this derivative also as an additional (extended) order parameter to characterize transition between layers. The dilation invariants are therefore:
\begin{equation}\label{eq:3.10}
{{\pmb{\rm{I}}}_1} = \ell _{}^*/{y^*}^\alpha  = \ell /{y^\alpha }
\end{equation}
 \begin{equation}\label{eq:3.11}
{{\pmb{\rm{I}}}_2} = \left( {\frac{{d\ell _{}^*}}{{d{y^*}}}} \right)/{y^*}^{(\alpha  - 1)} = \left( {\frac{{d\ell }}{{dy}}} \right)/{y^{(\alpha  - 1)}}
 \end{equation}
Below, we will introduce three kinds of invariant solutions (note that they are not the only invariant solutions - as any $F$ in \eqref{eq:3.5} is a candidate), while the reasons why they exist and how they agree with DNS data will be presented in the next section.

\subsection{Power law}
The first kind is constant dilation invariant for $\pmb{\rm{I}}_1$, which leads to the power law scaling for the stress length function:
\begin{equation}\label{eq:3.12}
\text{If \quad \quad } \pmb{\rm{I}}_1 = \text{const.}, \text{\quad then \quad }  \ell =\pmb{\rm{I}}_1 y^{\alpha}
\end{equation}
In this case, $\pmb{\rm{I}}_2$ is also constant, which can be verified as:
\begin{equation}\label{eq:3.13}
{{\pmb{\rm{I}}}_2}{\rm{ = }}\alpha {{\pmb{\rm{I}}}_1}{\rm{ = const}}..
\end{equation}
The ratio of the two dilation invariants is particularly important:
\begin{equation}\label{eq:3.14}
\gamma  = {{\pmb{\rm{I}}}_2}/{{\pmb{\rm{I}}}_1} = dln(\ell )/dln(y)
\end{equation}
We suggest to use $\gamma$ as a diagnostic function to analyze DNS data: if $\gamma$ displays a plateau (constant $\alpha$) in a range of $y$, then there is a local power law of $\ell$. How $\gamma$ enables discovery of different scaling exponents in the viscous sublayer, buffer layer, etc., will be shown later in figure 1.

\subsection{Defect power law}
When the dilation symmetry for $\ell$ is broken (${{\pmb{\rm{I}}}_1} \ne {\rm{const}}{\rm{.}}$), we propose a second kind of invariant relations:
\begin{equation}\label{eq:3.15}
\text{For \quad \quad }{{\pmb{\rm{I}}}_1} \ne {\rm{const}}{\rm{.}},\text{\quad but \quad } {{\pmb{\rm{I}}}_2}{\rm{ = const}}{\rm{.}} \text{\quad we have \quad } d\ell /dy = {{\pmb{\rm{I}}}_2}{y^{(\alpha  - 1)}}
\end{equation}
Integration of \eqref{eq:3.15} yields the defect power law:
\begin{equation}\label{eq:3.16}
\ell  = ({{\pmb{\rm{I}}}_2}/\alpha ){y^\alpha } + c{\rm{ }}
\end{equation}

\subsection{Scaling transition - a generalized invariant relation}

{To describe a smooth and monotonic transition of $\ell$ from one scaling $\ell ^{(I)} = c_I^{}{y^{{\gamma _I}}}$ to another $\ell ^{(II)} = c_{II}^{}{y^{{\gamma _{II}}}}$}, a simple transition ansatz can be found with a nonlinear relation between the two dilation invariants. Specifically, we extend (3.11) to include a nonlinear term as
\begin{equation}\label{eq:3.17}
{{\pmb{\rm{I}}}_2} = {\gamma _I}{{\pmb{\rm{I}}}_1} + c{\left( {{{\pmb{\rm{I}}}_1}} \right)^n} \quad \Rightarrow \quad \ell  = c_I{y^{{\gamma _I}}}{\left( {1 + {{(y/{y_c})}^p}} \right)^{({\gamma _{II}} - {\gamma _I})/p}}{\rm{ }}
\end{equation}
{where ${{\pmb{\rm{I}}}_1}=\ell/y^{\gamma_{II}}$ and $ {{\pmb{\rm{I}}}_2} =({d\ell/dy})/y^{\gamma_{II}-1} $} (algebraic calculation shown later). The two adjacent power laws are
\begin{equation}\label{eq:3.18}
{\ell ^{(I)}} = c_I{y^{{\gamma _I}}} \text{\quad for \quad } y \ll y_c
\end{equation}
\begin{equation}\label{eq:3.19}
{\ell ^{(II)}} = (c_Iy_c^{{\gamma _I} - {\gamma _{II}}}){y^{{\gamma _{II}}}} \text{\quad for \quad }  y \gg y_c
\end{equation}
and the transition location $y=y_c={\left( {c_I^{}/c_{II}^{}} \right)^{1/({\gamma _{II}} - {\gamma _I})}}$ is the cross point where $
{\ell ^{(I)}} = {\ell ^{(II)}} $. Note that \eqref{eq:3.17} works as an interpolation function in \citet{batchelor1951} for velocity structure function, and in a model for MVP by \citet{lvovprl}.

The choice of the generalized invariant relation \eqref{eq:3.17} is explained below. {Note that the left hand side (l.h.s.) of \eqref{eq:3.17} can be rewritten as
${{\pmb{\rm{I}}}_2}/{{\pmb{\rm{I}}}_1}=\gamma_I+c{\left( {{{\pmb{\rm{I}}}_1}} \right)^{n-1}}$, characterizing the local scaling exponent $\gamma={{\pmb{\rm{I}}}_2}/{{\pmb{\rm{I}}}_1}$ varying from a constant value $\gamma_I$ (in layer I) to $\gamma_{II}$ (in layer II) by proper $c$ and $n$. To see this, we further rewrite the l.h.s. of \eqref{eq:3.17} as:
\begin{equation}\label{eq:3.20}
\frac{{\gamma  - {\gamma _I}}}{{{\gamma _{II}} - {\gamma _I}}} = {\left( {\frac{{{c_{II}}}}{{{\pmb{\rm{I}}}_1{}}}} \right)^q}
\end{equation}
where $n=1-q$ and $c=({\gamma _{II}} - {\gamma _I} )c_{II} ^q$ are substituted in}. Then, for $y \gg y_c$ (approaching layer II), ${\pmb{\rm{I}}}_1=\ell/y^{\gamma_{II}} \to c_{II}$, the right hand side (r.h.s.) of \eqref{eq:3.20} goes to 1, as expected for the l.h.s. of \eqref{eq:3.20}. For $y \ll y_c$ (approaching layer I), the r.h.s. of \eqref{eq:3.20} goes to zero when $p = q({\gamma _{II}} - {\gamma _I}) \gg 1$, also as expected for the l.h.s. of \eqref{eq:3.20}. This can always be guaranteed by choosing an appropriate $q$ (thus $q$ specifies the sharpness of the transition between the two layers). Therefore, \eqref{eq:3.20} connecting the two asymptotic power law states is explicitly written as below
\begin{equation}\label{eq:3.21}
\frac{{d{{(\ell /{y^{{\gamma _I}}})}^{p/({\gamma _{II}} - {\gamma _I})}}}}{{d({y^p})}} = c_{II}^{p/({\gamma _{II}} - {\gamma _I})}
\end{equation}
which yields the scaling transition ansatz - the r.h.s. of \eqref{eq:3.17} - after integration:
\begin{equation}\label{eq:3.22}
\ell  = {c_I}{y^{{\gamma _I}}}{\left( {1 + {{(y/{y_c})}^p}} \right)^{({\gamma _{II}} - {\gamma _I})/p}}{\rm{ }}
\end{equation}
All of the parameters in \eqref{eq:3.22} are determined from \eqref{eq:3.20}, except for $c_\textit{I}$ which is an integration constant determined by the power law coefficient in layer I.

In the section below, we will show that the power law \eqref{eq:3.12} describes the stress length function in the viscous sublayer, buffer layer, core layer; the defect power law \eqref{eq:3.15} describes the bulk flow - also see \cite{wuyoups}; and the scaling transition ansatz \eqref{eq:3.17} connects two adjacent layers together. Their combination yields a complete formula for the stress length function over the entire flow domain.

\section{A multi-layer stress length function }

In this section, a multi-layer formula for the entire profile of stress length function is presented. Note that the notion of the multi-layer is well known in the literature \cite{pope2000turbulent}, i.e. viscous sublayer, buffer layer, log law region (overlap region), etc., but has not received a complete quantitative description yet. According to our study, a bulk flow region can be defined by quasi-balance between production and dissipation. This bulk flow is connected to the overlap region near the wall, and extends, away from the wall, to the edge of TBL, but ends at the `core layer' in channel and pipe where turbulent transport replaces production to balance dissipation. This notion of using different leading balance of turbulent kinetic energy budget to characterize different layers is new, and it is also novel to relate them to different (dilation) symmetries. Since the stress length (order) function reveals different symmetries in these layers with different local scaling laws (i.e. power law and defect power law), it is qualified to be an order function.

It is necessary to note that we introduce inner and outer scales to normalize the stress length function (and balance equations) and to characterize $\emph{Re}$ scaling of MVP. Specifically, viscous units ($y^+$ and $u_\tau$) are used for the inner flow. For the outer flow, velocity scale is still $u_\tau$, while the length scale is different. We define an outer dilation center (denoted as $y=\delta$), i.e. the centerline of channel/pipe, or a new edge of TBL ($\delta$, which is not necessarily $\delta_{99}$ commonly used to describe the boundary layer edge of TBL). Concretely, $r=1-y/\delta$ defines the normalized distance to the outer dilation center ($r=0$ or $y=\delta$) and is an independent variable to take place of wall distance $y$. The invariant solutions expressed in terms of the stress length function are the same as \eqref{eq:3.12}, \eqref{eq:3.15} and \eqref{eq:3.17}, only by replacing $y$ with $y^+$ (inner) or with $r$ (outer). At the end, those postulated invariant solutions should be confirmed by DNS data, as we present in detail as below.

\subsection{Inner flow}
Using viscous units, i.e.
\begin{equation}\label{eq:4.1}
{y^ + } = {y u_\tau}/{{\nu { }}},\quad {U^ + } = {U}/{{{u_\tau }}}
\end{equation}
the streamwise mean momentum equation (taking channel flow as an example), is thus:
\begin{equation}\label{eq:4.2}
{C} = \frac{{{\partial ^2}{U^ + }}}{{\partial {y^{ + 2}}}} + 2\ell _M^{ + 2}\left( {\frac{{\partial {U^ + }}}{{\partial {y^ + }}}} \right)\left( {\frac{{{\partial ^2}{U^ + }}}{{\partial {y^{ + 2}}}}} \right) + 2\ell _M^ + \dot \ell _M^ + {\left( {\frac{{\partial {U^ + }}}{{\partial {y^ + }}}} \right)^2} + \frac{1}{{R{e_\tau }}} = 0
\end{equation}
where $\dot \ell _M^ +  = \partial \ell _M^ + /\partial {y^ + }$ is the derivative of stress length function. It admits
following dilation transformations according to \eqref{eq:3.3},
\begin{equation}\label{eq:4.3}
{y^{ + *}} = {e^{\epsilon}}{y^ + },{\rm{    }}Re_\tau ^* = {e^{(1 + 2\alpha )\epsilon}}R{e_\tau },{\rm{    }}{U^{ + *}} = {e^{(1 - 2\alpha )\epsilon}}{U^ + },{\rm{    }}\ell _M^{ + *} = {e^{\alpha \epsilon}}\ell _M^ +
\end{equation}
The normalized group invariants for stress length function \eqref{eq:3.10} and its derivative \eqref{eq:3.11}, are respectively:
\begin{equation}\label{eq:4.4}
{{\pmb{\rm{I}}}_1} = \ell _M^ + /{y^{ + \alpha }},\quad \quad \quad {{\pmb{\rm{I}}}_2} = \dot \ell _M^ + /{y^{ + (\alpha  - 1)}}
\end{equation}
Then, the constant dilation invariant assumption in (3.12) is:
\begin{equation}\label{eq:4.5}
{{\pmb{\rm{I}}}_1} = {c_1},\quad and\quad {{\pmb{\rm{I}}}_2} = \alpha {c_1}
\end{equation}
which yields a power law scaling as a function of $y^+$:
\begin{equation}\label{eq:4.6}
\ell _M^ +  = {c_1}{y^{ + \alpha }}
\end{equation}

To test \eqref{eq:4.6}, it is suggested to display the following diagnostic function as in \eqref{eq:3.14}:
\begin{equation}\label{eq:4.7}
\gamma  = {{\pmb{\rm{I}}}_2}/{{\pmb{\rm{I}}}_1} = dln(\ell _M^ + )/dln({y^ + })
\end{equation}
If the empirical $\gamma$ displays a plateau in a range of $y^+$, then, a local power law of $\ell _M$ is validated, and the value of the plateau is thus $\alpha$. This is shown in figure 1, with $\alpha$=3/2 in the viscous sublayer, $\alpha$=2 in the buffer layer, and $\alpha$=1 in the log layer. Note that in order to present a clear display of $\alpha$=1 in the log layer, we plot a compensated $\gamma$ function, i.e. $
Q({y^ + }) = dln(\ell _M^{ + DNS}/L)/dln({y^ + })$, where
$L = \ell _M^{Outer}/y^+ $ is the theoretical formula for the outer flow (see later herein). This compensated plot eliminates the outer flow influence on the log layer, but without changing the scaling exponent in the viscous sublayer and buffer layer, as $L$ is constant near wall. Below we will introduce the local power law for each of the layers.

\begin{figure}
  \includegraphics[width=7cm]{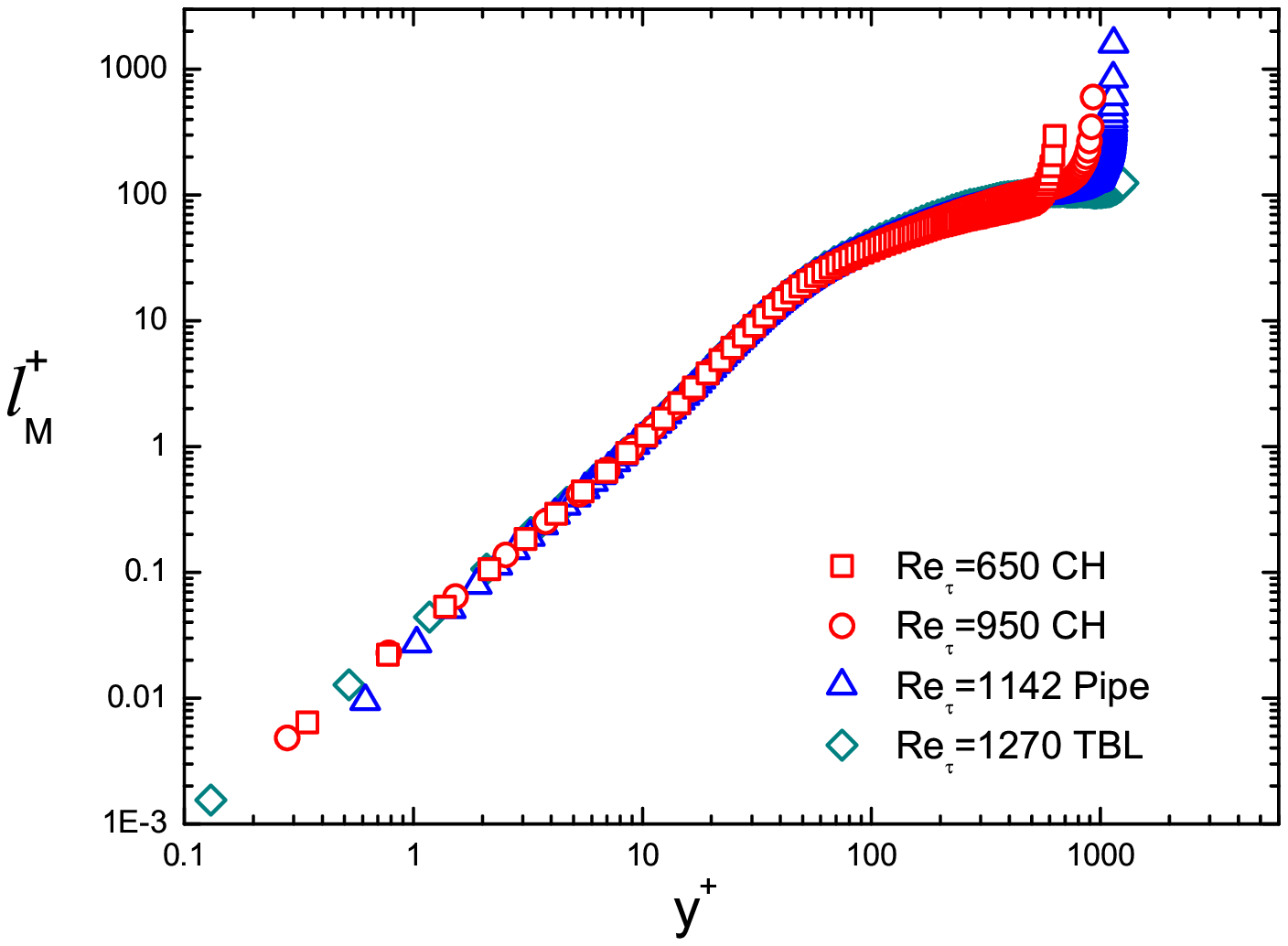}
  \includegraphics[width=7cm]{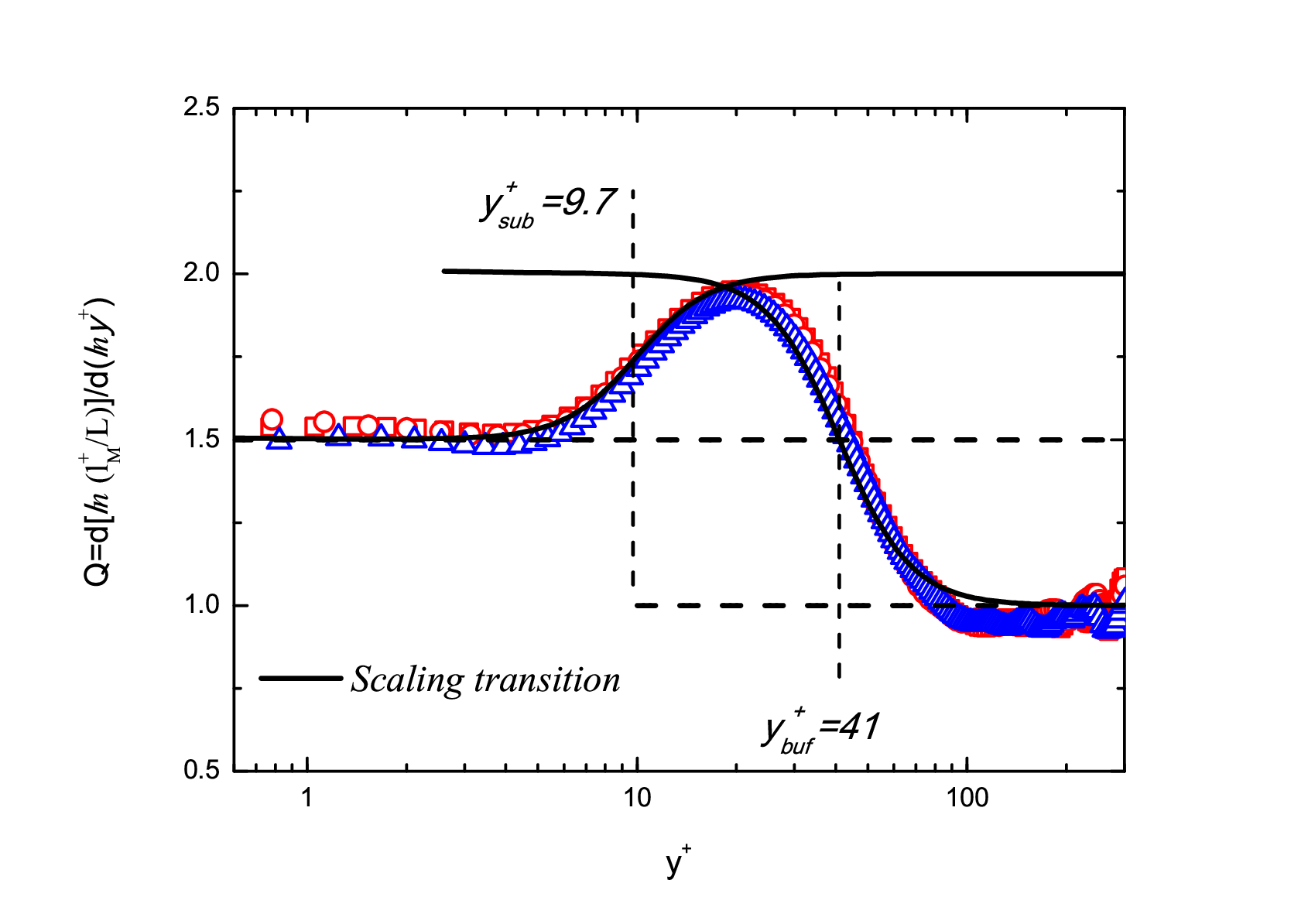}
\caption{[left] stress length function shown by DNS data. [right]: $Q({y^+})$ reveals local scaling in sub, buffer, and log layers with exponents 3/2, 2, and 1, respectively. Two channel flows from \cite{Iwamoto2002} at $Re_{\tau}=650$ and \citet{hoyas2006} at $Re_{\tau}=940$, one pipe flow of \citet{wuxiaohua} at $Re_{\tau}=1142$, and one TBL flow of \citet{schlatter2010simulations} at $Re_{\tau}=1270$. Dashed lines indicate sublayer thickness $y^{+}_{sub}=9.7$ and buffer layer thickness $y^{+}_{buf}=41$, respectively, at the middle of scaling transitions.}
\label{fig:1}
\end{figure}

\subsubsection{Viscous Sublayer}
In the viscous sublayer, the stress length function is:
 \begin{equation}\label{eq:4.8}
\ell _M^{ + sub} = {\pmb{\rm{I}}}_1^{sub}{y^{ + 3/2}}
 \end{equation}
This can be justified by a near-wall expansion \cite{wuyouscp}: $u' \propto y$ and $\nu '\propto y^2 $ such that $W^+ \propto y^{+3}$. Since $S^{+} \approx 1$ near the wall, thus $\ell _M^ +  \propto {y^{ + 3/2}}$ (hence $\alpha=3/2$). Note that for ${\pmb{\rm{I}}}_1^{sub}$, according to current DNS data, it is about 0.034.

\subsubsection{Buffer layer}

Note that the log-layer is well known as in \eqref{eq:lmoverlap}, while the viscous sublayer is characterized by \eqref{eq:4.8}. A natural question is: what is the scaling exponent of the stress length function in the buffer layer? In fact, in the buffer-layer, the power law for the stress length function is:
\begin{equation}\label{eq:4.9}
\ell _M^{ + buf} = {\pmb{\rm{I}}}_1^{buf}{y^{ + 2}}
\end{equation}
Preliminary study yields the following explanation. Using dimensional analysis, we obtain ${\ell _M} = {\ell _\nu }{\Theta ^{1/4}}$, where ${\ell _\nu } = {(W/S)^{3/4}}/{{\rm{\varepsilon }}^{1/4}} = \nu _T^{3/4}/{{\rm{\varepsilon }}^{1/4}} $ is a shear-induced eddy length, $\Theta=\varepsilon /SW$ is the ratio between turbulent dissipation and production. A near-wall expansion yields ${\ell _\nu } \propto {y^2}$, while in the buffer layer
$\Theta  \approx const$, due to the fact that turbulent transport reaches a local maximum and dissipation is of the same order as production. Hence, ${\ell _M} \propto {y^2}$ by multiplying $\ell _{\nu}$ and $\Theta$. Such a scaling exponent 2 is shown in figure 1, indicated by the peak of $\gamma$ function located at about $y^+ =20$; and the coefficient ${\pmb{\rm{I}}}_1^{buf} \approx 0.01 $ for moderate $Re$'s. Whether this power law can be explained by a statistical study of coherent vortex structures \cite{schoppa2002coherent} in the buffer layer, deserves further study. Interestingly, \eqref{eq:lmsublayer} assumes also a power law scaling with exponent 2, but it should be valid in the buffer layer, not viscous sublayer.

\subsubsection{Log law region (log layer)}
The power scaling in the log law region (log layer) is
\begin{equation}\label{eq:4.10}
\ell _M^{ + log} = {\pmb{\rm{I}}}_1^{log}{y^ + } = \kappa {y^ + }
\end{equation}
which is the classical assumption made by Prandtl in 1925 (leading to the log law for the mean velocity), i.e. the Karman constant $\kappa  = {\pmb{\rm{I}}}_1^{log}$. Later we will see \eqref{eq:4.10} can be obtained from a near wall asymptotic state of a (outer) bulk solution.

\subsection{Scaling laws in the outer flow}
Normalization using outer length scale ($\delta$) is
 \begin{equation} \label{eq:4.11}
        r=1-y/\delta,\quad \ell _M ^{\wedge}  = {\ell _M}/\delta
 \end{equation}
which $\delta$ is the half height of channel or pipe radius; while for TBL, $\delta$ is not necessarily the boundary layer edge, hence denoted as $\delta=\sigma \delta_{99}$, where $\sigma$ is a coefficient close to one. Then, the mean momentum equation \eqref{eq:4.2} is:
  \begin{equation}\label{eq:4.12}
 {N} = \frac{- 1}{Re_{\tau }}\frac{\partial ^2 U^ + }{\partial r^2} + 2\ell _M^{\wedge 2}\frac{\partial U^ + }{\partial r}\frac{\partial ^2 U^+ }{\partial r^2} + 2\ell _M^{\wedge} \dot {\ell} _M^{\wedge}\left( \frac{\partial U^ + }{\partial r} \right)^2 - 1 = 0
  \end{equation}
where $\dot{ \ell} _M^{\wedge} = \partial \ell _M^{\wedge}/\partial r $. The group invariants for the length function and its derivative are respectively
\begin{equation}\label{eq:4.13}
    {\pmb{\rm{I}}}_1 = \ell _M^{\wedge}/r^{\alpha},\quad \quad \quad {\pmb{\rm{I}}}_2 = \dot \ell _M^{\wedge}/r^{\alpha  - 1}
\end{equation}
and the corresponding diagnostic function for the power law scaling exponent is:
\begin{equation}\label{eq:4.14}
\gamma  = {\pmb{\rm{I}}}_2/{\pmb{\rm{I}}}_1 = \partial ln(\ell _M^{\wedge})/\partial ln(r)
\end{equation}

\subsubsection{Core layer}

A power law is obtained near the centerline (core layer), where the stress length function diverges as:
\begin{equation}\label{eq:4.15}
\ell _M^{\wedge core} = {\pmb{\rm{I}}}_1^{core}r^{- 1/2}
\end{equation}
Note that as $r \rightarrow 0$, $W^+ \approx r$ and $S^+ \propto r$ (central symmetry), hence, $\ell _{M}^{\wedge} \propto r^{-1/2}$. This extra layer is present in a channel/pipe, as shown in figure 2. For TBL, the core layer is absent, since no center symmetry is forced by opposite wall condition (see figure 3). Note that according to DNS data, out of the boundary layer edge of TBL, the stress length function also increases quickly, as that in core layer for channel and pipe. However, this region covers a very small percentage in the variation of the streamwise mean velocity less than 1\%, hence ignored presently.

\subsubsection{Bulk flow (Quasi-balance region)}
The bulk flow is defined by a quasi-balance between production ($SW$) and dissipation ($\varepsilon$). In this region, $\Theta =\varepsilon/SW \approx 1$, and $\ell_{\nu}=\nu^{3/4}_{T}/\varepsilon ^{1/4} \rightarrow \ell_{0}$ as $r \rightarrow 0$ (finite dissipation and eddy viscosity), therefore $\ell _{M}=\ell _{\nu}\Theta ^{1/4} \approx \ell_{\nu} \rightarrow \ell_{0}$. The existence of a finite $\ell _{0}$ as a characteristic length scale breaks dilation symmetry of $\ell _{M}$, and the constant group invariant assumption of the length function \eqref{eq:3.12} is not valid in the bulk flow. That is why we introduce the constant group invariant for its first derivative, i.e.
\begin{equation}\label{eq:4.16}
{{\pmb{\rm{I}}}_1} \ne {\rm{const}}{\rm{.}},\quad and\quad {{\pmb{\rm{I}}}_2}{\rm{ = const}}{\rm{.}}
\end{equation}
In other words, although dilation symmetry is broken in the length function, we assume it is preserved in its derivative. Integrating \eqref{eq:4.16} using definitions in \eqref{eq:4.13} yields
\begin{equation}\label{eq:4.17}
\ell _M^{\wedge} = ({{\pmb{\rm{I}}}_2}/m){r^m} + c{\rm{ }}
\end{equation}
where $c$ is an integration constant and $m$ (=$\alpha$) is an scaling exponent.

Note that \eqref{eq:4.17} should be consistent with the wall condition:$\ell _M^{\wedge} \to 0$ as $r \rightarrow 1$ (towards the wall), then $c{\rm{  = }} - {{\pmb{\rm{I}}}_2}/m$. A consequence is that we obtain a linear asymptotic scaling for $\ell _M^{\wedge}$ as $r \rightarrow 1$, i.e. $\ell _M^{\wedge} \to  - {{\pmb{\rm{I}}}_2}(1 - r) =  - {{\pmb{\rm{I}}}_2}(y/\delta )$, which is, in viscous units, $\ell _M^ +  \to  - {{\pmb{\rm{I}}}_2}{y^ + }$. This is exactly the linear scaling (4.10) in the log layer. If we define $\kappa  =  - {{\pmb{\rm{I}}}_2}$, then \eqref{eq:4.17} matches exactly with (4.10) when  $r \rightarrow 1$. Therefore, final expression for the bulk flow is
\begin{equation}\label{eq:4.18}
\ell _M^{\wedge bulk} = \kappa (1 - {r^m})/m
\end{equation}
Note that DNS data (figure 3) suggest that m=4 for channel and TBL, and m=5 for pipe; a derivation following Landau's mean-field argument is presented later herein.

\subsection{Scaling transition between adjacent layers}

Above results give a (local) quantitative characterization of the multi-layer structure, i.e. power law for viscous sublayer, buffer layer, core layer and log layer, and defect power law for bulk flow. As proposed by \citet{Oberlack2010new}, it is an open issue to describe scaling matching between different layers. Now let us address this issue by using the generalized invariant relation \eqref{eq:3.17} as below.

The generalized invariant relation in (3.17) leads to a scaling transition ansatz \eqref{eq:3.20}, smoothly describes the transition from one layer to another. We use \eqref{eq:3.20} for the viscous sublayer and buffer layer, i.e.
\begin{equation}\label{eq:4.19}
\frac{{\gamma  - 3/2}}{{2 - 3/2}} = {\left( {\frac{{{\pmb{\rm{I}}}_1^{buf}}}{{\ell _M^ + /{y^{ + 2}}}}} \right)^{{q_1}}}
\end{equation}
where 3/2 and 2 are the scaling exponents in the sublayer and buffer layer, respectively; and ${\pmb{\rm{I}}}_1^{buf}$ is the constant dilation invariant in the buffer layer. It leads to the following scaling transition connecting the two layers - \eqref{eq:4.8} and \eqref{eq:4.9} - together:
\begin{equation}\label{eq:4.20}
\ell _M^{ + (sub - buf)} = {\pmb{\rm{I}}}_1^{sub}{y^{ + 3/2}}{\left( {1 + {{\left( {\frac{{{y^ + }}}{{y_{sub}^ + }}} \right)}^{{p_1}}}} \right)^{\frac{1}{{2{p_1}}}}} = \rho {\left( {\frac{{{y^ + }}}{{y_{sub}^ + }}} \right)^{\frac{3}{2}}}{\left( {1 + {{\left( {\frac{{{y^ + }}}{{y_{sub}^ + }}} \right)}^{{p_1}}}} \right)^{\frac{1}{{2{p_1}}}}}
\end{equation}
where the constant dilation invariants ${\pmb{\rm{I}}}_1^{sub}$ and ${\pmb{\rm{I}}}_1^{buf}$ are taken placed by $y_{sub}^ += ({\pmb{\rm{I}}}_1^{sub}/{\pmb{\rm{I}}}_1^{buf})^2$ (which is called the sublayer thickness, and takes a value of about 9.7 derived from DNS data), $\rho  = {\pmb{\rm{I}}}_1^{sub}{{\rm{(I}}_1^{sub}/{\pmb{\rm{I}}}_1^{buf})^3}\approx 1.03$ (also determined from DNS data). Note that the transition sharpness $p_{1}=q_1(2-3/2)=q_1/2$ is set as integer 4, which is least sensitive to predict MVP (and can be set as 6 for example).

Similarly, for buffer and log layers, by using \eqref{eq:3.20} :
\begin{equation}\label{eq:4.21}
\frac{{\gamma  - 2}}{{1 - 2}} = {\left( {\frac{{{\pmb{\rm{I}}}_1^{log}}}{{\ell _M^ + /{y^ + }}}} \right)^{{q_2}}}
\end{equation}
which leads to the following scaling transition:
\begin{equation}\label{eq:4.22}
\ell _M^{ + (buf - log)} = {\pmb{\rm{I}}}_1^{buf}y_{}^{ + 2}{\left( {1 + {{\left( {\frac{{{y^ + }}}{{y_{buf}^ + }}} \right)}^{{p_2}}}} \right)^{\frac{{ - 1}}{{{p_2}}}}} = \rho {\left( {\frac{{{y^ + }}}{{y_{buf}^ + }}} \right)^2}{\left( {1 + {{\left( {\frac{{{y^ + }}}{{y_{buf}^ + }}} \right)}^{{p_2}}}} \right)^{\frac{{ - 1}}{{{p_2}}}}}
\end{equation}
where $y_{buf}^ + {\rm{ = I}}_1^{\log }/{\pmb{\rm{I}}}_1^{buf} = \kappa y_{sub}^{ + 2}/\rho $ (about 41 for moderate $Re$'s from DNS data), and $p_2$ is also set as 4 as $p_1$.

The generalized invariant relation also works for the outer flow. Note that the bulk solution \eqref{eq:4.18} saturates to a constant $\ell _M^{\wedge} \to \kappa /m$, indicating a scaling exponent zero; while the scaling exponent in the core layer is -1/2, thus \eqref{eq:3.20} connecting the bulk edge and the core layer is
\begin{equation}\label{eq:4.23}
\frac{{\gamma  - 0}}{{ - 1/2 - 0}} = {\left( {\frac{{{\pmb{\rm{I}}}_1^{core}}}{{\ell _M^{\wedge}\sqrt r }}} \right)^{{q_3}}}
\end{equation}
where $\gamma$ is a function of $r$ as in \eqref{eq:4.14}. It leads the following composite solution
\begin{equation}\label{eq:4.24}
\ell _M^{(bulk - core)} = (\kappa /m){(1 + {(r/{r_{core}})^{{p_3}}})^{ - 1/(2{p_3})}}/{Z_c}
\end{equation}
where ${Z_c} = {(1 + {(1/{r_{core}})^{{p_3}}})^{ - 1/(2{p_3})}}$ is integrated coefficient to guarantee $
\ell _M^{\wedge} \to \kappa /m$ as $r \to 1$, and ${\pmb{\rm{I}}}_1^{core} = (\kappa r_{core}^{1/2})/(m{Z_c})$  (where ${\pmb{\rm{I}}}_1^{core} \approx 0.058$) is taken placed by the core layer thickness $r_{core}$ (which is about 0.27 for current DNS data). Note that the sharpness $p_{3}$ here can be derived, to be -2 from a central symmetry consideration, as presented in the next section.

Let us summarize above invariant solutions in the following Table I. Note that the defect power law in the quasi-balance region connects two asymptotic scalings, i.e. linear scaling in the log layer, and finite value at bulk edge. Also note that for TBL, there is no core layer (see Part IV). These local scalings are also shown in figure 2.

\begin{figure}\label{tab:1}
\centering
\includegraphics[width=14cm]{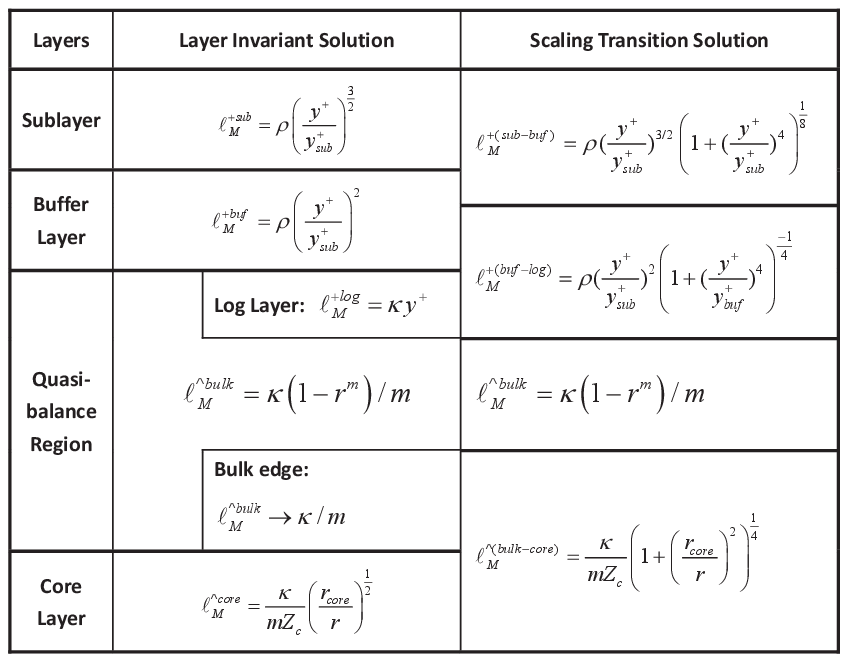}
\caption*{Table I. Multi-layer solution for the stress length
function. The right column shows the scaling transition solution
which connects local power laws in adjacent layers (middle column)
together. Note that $m=4$ for channel and TBL, $m=5$ for pipe, and
$\rho  = \kappa y_{sub}^{ + 2}/y_{buf}^ + $, ${Z_c} = {(1 +
r_{core}^2)^{1/4}}$, while there are four parameters, i.e.
$y_{sub} ^+$, $y_{buf} ^ +$, $r_{core}$ and $\kappa$ to be determined in Part II \& III.}
\end{figure}

\begin{figure}
\centering
\includegraphics[width=12cm]{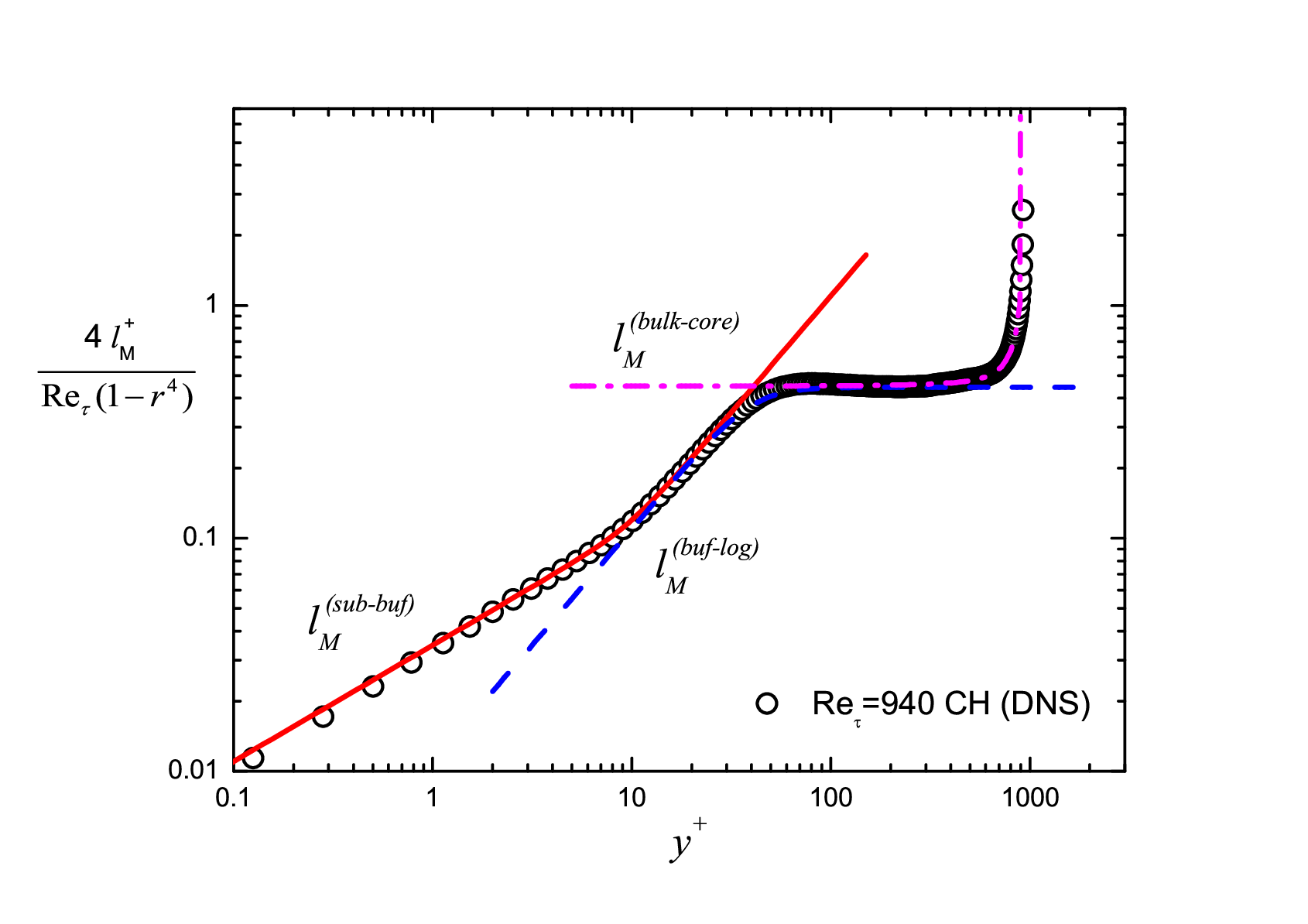}
\caption{ A compensated plot of the stress length function (symbols) divided by the bulk solution (e.g. $1-r^4$). The plateau in the bulk flow indicates the defect power law (in coordinate $r$) from $y^+$ at about 50 to about 0.6$Re_{\tau}$. The lines are composite solutions: the sub-buffer layer transition $\ell _M^{ + (sub - buf)}$  (compensated by $y^+$, solid line), the buffer-log layer transition $\ell _M^{ + (buf - \log )}$ (compensated by $y^+$, dashed line), and the bulk-core layer transition $ \ell _M^{ + (bulk - core)}$ (dashed line). See Table I.}
\label{fig:2}
\end{figure}

\subsection{Composite solution of stress length function for the entire flow}
To obtain a composite solution in the entire flow domain, we use the following multiplicative rule \cite{van1964},
\begin{equation}\label{eq:4.25}
{\phi ^{I - III}} = {\phi ^{I - II}}{\phi ^{II - III}}/{\phi ^{Common}}
\end{equation}
Note that for the inner three layers, the multiplicative rule corresponds to
\begin{equation}\label{eq:4.26}
\ell _M^{ + In} = \ell _M^{ + (sub - buf)}\ell _M^{ + (buf - log)}/\ell _M^{ + buf}
\end{equation}
which leads to the following composite solution for the inner flow
\begin{equation}\label{eq:4.27}
\ell _M^{ + In} = \rho {\left( {\frac{{{y^ + }}}{{y_{sub}^ + }}} \right)^{\frac{3}{2}}}{\left( {1 + {{(\frac{{{y^ + }}}{{y_{sub}^ + }})}^4}} \right)^{\frac{1}{8}}}{\left( {1 + {{(\frac{{{y^ + }}}{{y_{buf}^ + }})}^4}} \right)^{\frac{{ - 1}}{4}}}
\end{equation}
Similarly, applying the multiplicative rule to outer flow, i.e. \eqref{eq:4.18} and \eqref{eq:4.24},
\begin{equation}\label{eq:4.28}
\ell _M^{\wedge Outer} = \ell _M^{\wedge bulk}\ell _M^{\wedge (bulk - core)}/\ell _0^{\wedge}
\end{equation}
where $\ell _0^{\wedge} = \kappa /m$ is the common state, and the resulted outer solution is
\begin{equation}\label{eq:4.29}
\text{CH \& Pipe:} \quad \ell _M^{\wedge Outer} = \frac{\kappa }{m{Z_c}}(1 - r^m)\left( {1 + (\frac{r_{core}}{r})}^2 \right)^{\frac{1}{4}}
\end{equation}
\begin{equation}\label{eq:4.30}
\text{TBL: } \quad \ell _M^{\wedge Outer} = \frac{\kappa }{4}(1-r^4)
\end{equation}

A composite solution for the stress length function for the entire flow domain is obtained by applying the multiplication rule:
\begin{equation}\label{eq:4.31}
\ell _M^ +  = \ell _M^{ + In}\ell _M^{ + Outer}/\ell _M^{ + Common} = \ell _M^{ + In}\ell _M^{ + Outer}/\ell _M^{ + log}
\end{equation}
which is (for channel and pipe)
\begin{equation}\label{eq:4.32}
\ell _M^ +  = \rho \left( \frac{y^ + }{y_{sub}^ + } \right)^{\frac{3}{2}}{\left( 1 + {(\frac{y^ + }{y_{sub}^ + })}^4 \right)^{\frac{1}{8}}}\left( 1 + {(\frac{y^ + }{y_{buf}^ + })}^4 \right)^{\frac{ - 1}{4}}\frac{1 - r^m}{m(1 - r)Z_c}\left( 1 + {(\frac{r_{core}}{r})}^2 \right)^{\frac{1}{4}}
\end{equation}
For TBL, the entire formula is the same, except for the absence of the core layer:
\begin{equation}\label{eq:4.33}
\ell _M^ +  = \rho \left( \frac{y^ + }{y_{sub}^ + } \right)^{\frac{3}{2}}\left( 1 + {(\frac{y^ + }{y_{sub}^ + })}^4 \right)^{\frac{1}{8}}\left( 1 + {(\frac{y^ + }{y_{buf}^ + })}^4 \right)^{\frac{ - 1}{4}}\frac{1 - r^4}{4(1 - r)}
\end{equation}
Figure 3 shows verifications of \eqref{eq:4.32} and \eqref{eq:4.33} compensated by the bulk flow structure, $1-r^m$, in which each of the inner layers and the core (divergent) layer are also visible. The thickness of each layer is an important parameter, sketched in figure 3, whose accurate determination will be discussed in Part II and III. Also note that the stress length function in channel differs from that in pipe, by a different bulk flow scaling exponent, i.e. $m=4$ versus $m=5$ ; while in TBL, the bulk flow structure ($m=4$) extends almost to the boundary layer edge, and the core layer is absent, owing to the absence of the central symmetry.
\begin{figure}\label{fig:3}
\centering
\includegraphics[width=12cm]{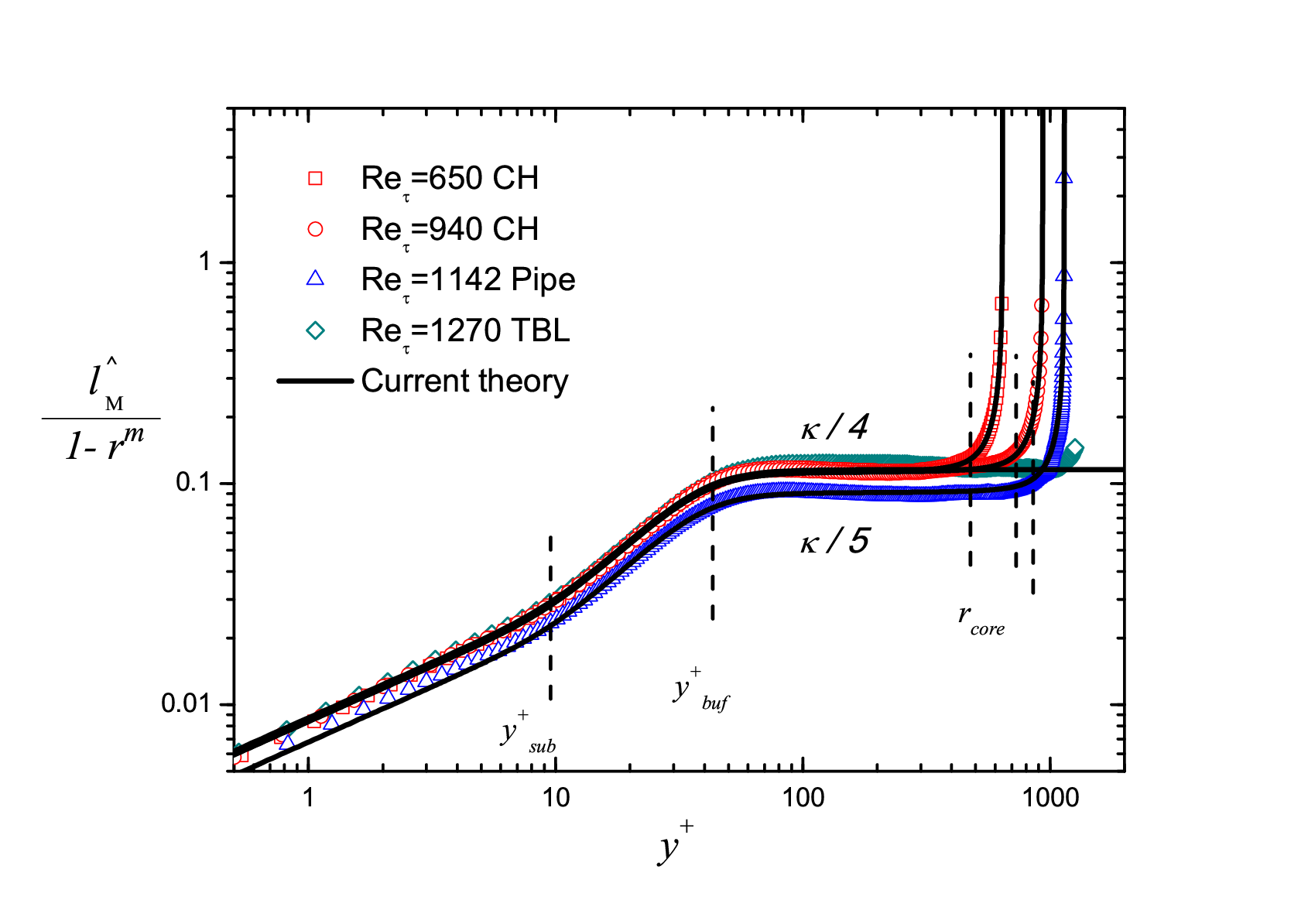} \\
\includegraphics[width=12cm]{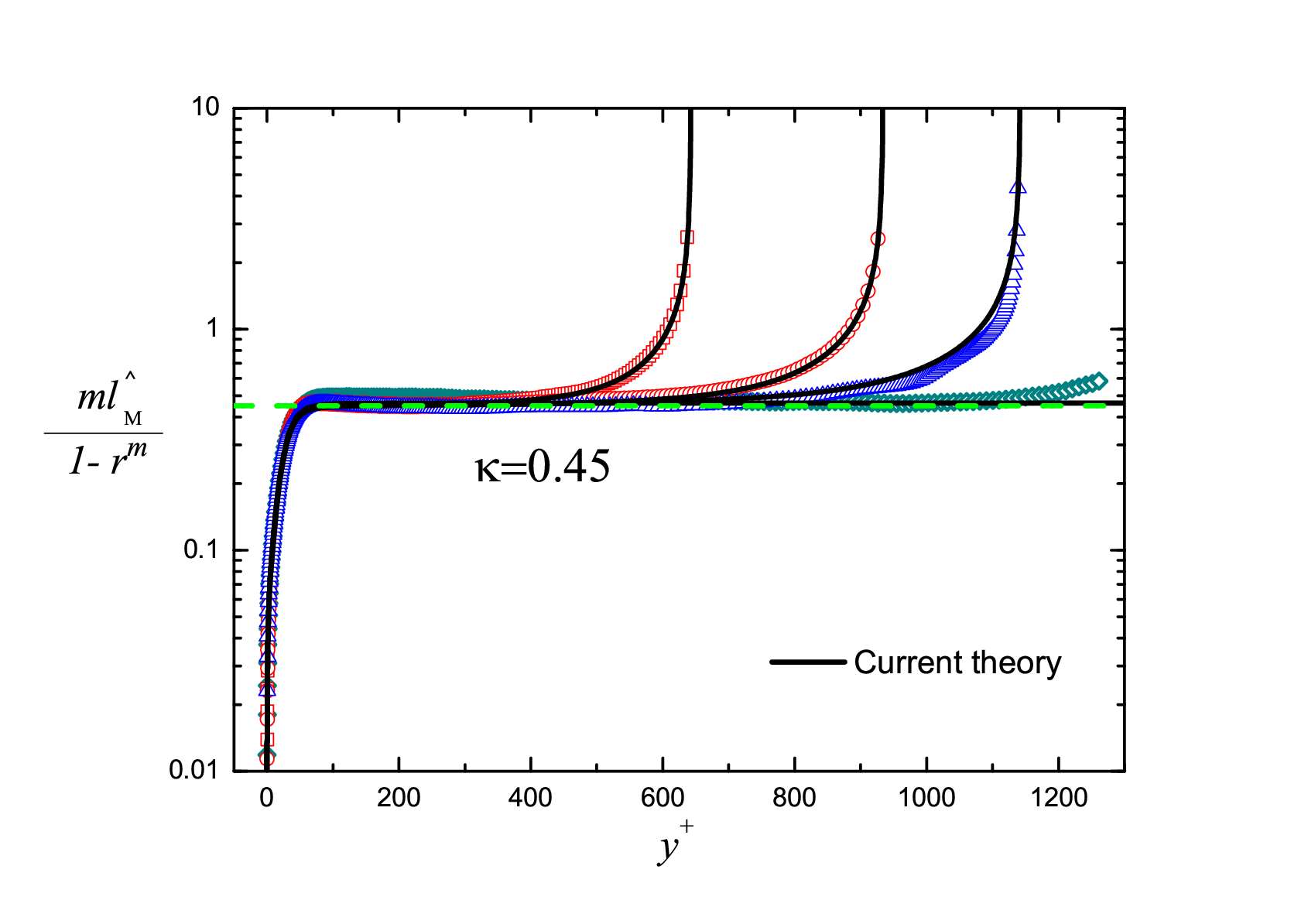}
\caption{ [top] The stress length function from DNS data (in log-log coordinate) divided by the bulk flow solution reveals a four-layer structure, i.e. viscous sublayer, buffer layer, bulk zone and core layer (for channel and pipe), separated by $y_{sub}^ +  \approx 9.7$, $y_{buf}^ +  \approx 41$ and ($r_{core} \approx 0.27$) respectively. [bottom] The stress length function in log-linear coordinate, to show the constant plateau indicating bulk flow structure, where the green dashed line indicates the bulk flow constant $\kappa =0.45$. Note that the stress length profiles in the three flows collapse in the viscous sublayer, buffer layer and bulk flow, by multiplying $m=4$ (channel and TBL) and $m=5$ (pipe) respectively. Solid lines are theoretical formulas (4.32) and (4.33) with above parameters (for current TBL, $\sigma \approx 0.7$). Also note that the plateau in TBL bulk flow is slightly larger than 0.45 at current low \emph{Re} (marginally visible); its $Re$-dependence will be discussed elsewhere.}
\end{figure}

\section{An interpretation for the bulk flow scaling}
One of the most interesting results described in the above sections is the bulk flow scaling, $1-r^m$, where $m=4$ for channel and $m=5$ for pipe. The two scaling exponents were initially obtained empirically from inspecting DNS data; here, we develop a theoretical argument which derives this scaling and explains the difference between channel and pipe. The interpretation is obtained by further exploring the analogy to Landau's mean field theory (Landau, 1958). With a similar variational calculation as that developed by Landau (1958), we obtain the functional form of the bulk stress length and clarify the relationship between the exponent $m$ and the geometry (flat plate versus circular wall). The analysis also considers energy budget for turbulent fluctuations, and predicts the profile of dissipation and transport in the outer flow.

Landau's mean field theory for ferromagnet \cite{Landau1958}
considered the phase transition near the critical temperature, by
assuming the thermodynamic potential density $\Omega$ expanded in
the order parameter, $\Psi (\textbf{r})$ - the magnetization at
position $\textbf{r}$, i.e.,
\begin{equation}\label{eq:5.1}
\Omega (\bf{r}) = {\Omega_0}(T) + h(\bf{r})\Psi (\bf{r}) + a(T)\Psi {(\bf{r})^2} + b(T)\Psi {(\bf{r})^4} + c(T){[\nabla \Psi (\bf{r})]^2} +  \cdots
\end{equation}
where $\Omega_0$ is the normal state potential before transition; $h(\textbf{r})$ is the external magnetic field; $a$, $b$, $c$ are material-dependent parameters (functions of temperature $T$). Consider the volume-integrated total free energy, $G = \int {\Omega dV}$, and minimize it with respect to the variation of $\Psi$, one obtains the Landau equation for describing the statistical equilibrium state:
\begin{equation}\label{eq:5.2}
h(\bf{r}) + 2a\Psi (\bf{r}) + 4b\Psi {(\bf{r})^3} - 2c\Delta \Psi (\bf{r}) = 0
\end{equation}
\eqref{eq:5.2} not only describes ferromagnetic phase transition, but also applies to other systems, such as superfluidity and superconductivity \cite{kadanoff2009more}. Although physical interpretations of $\Psi$ differ between systems, a general feature is that it connects one normal state indicated by $\Psi =0$, with another nonzero state indicated by $\Psi = \Psi _0$.

Since we regard $\ell _{M}^{+}$ as an order parameter (function) representative of the symmetry change, it is natural to develop a similar mean field equation for $\ell _{M}^{+}$. Let us first examine the known equations, the MME and mean kinetic energy equation (MKE), namely,
\begin{equation}\label{eq:5.3}
{S^ + } + {W^ + } = {\tau ^ + }
\end{equation}
\begin{equation}\label{eq:5.4}
{S^ + }{W^ + } + \Pi _p^ +  = {\varepsilon ^ + }
\end{equation}
where the $\Pi _p^ +$ is turbulent transport by pressure and normal velocity fluctuation, and $\varepsilon ^ + $ is the dissipation. From three quantities (mean shear $S$, Reynolds stress $W$ and dissipation $\varepsilon$), two length scales naturally arise from dimensional analysis:
\begin{equation}\label{eq:5.5}
\ell _M^{} = \sqrt {{W^{}}} /{S^{}} = \sqrt { - \left\langle {u'v'} \right\rangle } /(dU/dy)
\end{equation}
\begin{equation}\label{eq:5.6}
\ell _\nu = (\nu _T)^{3/4}/{\varepsilon ^{1/4}} = (W/{S^{}})^{3/4}/\varepsilon ^{1/4}
\end{equation}
{where $\ell _{\nu}$ denotes the shear-induced eddy length }- in contrast to the Kolmogorov dissipation length ($\nu ^{3/4}/\varepsilon ^{1/4}$). An exact relation between the two length functions is
\begin{equation}\label{eq:5.7}
\ell _M = \ell _\nu \Theta ^{1/4}
\end{equation}
where $\Theta  = \varepsilon /(SW)$ is the ratio of dissipation-production, and is also considered to be an order function of the second kind \citep{she2009}. Therefore, the stress length function is decomposed into two order functions, i.e. $\ell _{\nu}$ and $\Theta $, derived as below.

We assert that $\ell _{\nu}$ is the order parameter in the spirit of Landau, since it is zero at the wall and saturates to a finite value $\ell _0$ at the centerline, due to the fact that both eddy viscosity and dissipation tend to constants. Take the outer coordinate $r=1-y/\delta$, and use the super index $\wedge$ to indicate outer normalization, i.e. $\ell ^{\wedge} = \ell / \delta $, we propose a simple equation for $\ell_{\nu} ^{\wedge}$, similar to the Landau equation \eqref{eq:5.2}:
\begin{equation}\label{eq:5.8}
h({r}) + 2c\Delta \ell _\nu ^{\wedge}({r}) = 0
\end{equation}
where $h(r) = \int_0^r {rd\widehat \sigma } $ denotes the effective external force (e.g. an area integration for an effective stress linear with $r$ normalized with wall stress, inspired from total stress $\tau ^{+} =r$ in channel and pipe), and $ d\widehat \sigma $ is the normalized area element, i.e. $d\widehat \sigma =rdr$ for pipe, $d\widehat \sigma =dr$ for channel and TBL. Note that \eqref{eq:5.8} leads to
${r^3}/3 + c({d^2}\ell _\nu ^{\wedge}/d{r^2}) = 0$ for pipe, and  ${r^3}/4 + c({d^2}\ell _\nu ^{\wedge}/d{r^2}) = 0$ for channel and TBL. Thus, with boundary conditions $d{\ell _\nu }/dr = 0$ at $r=0$ , and $\ell _{\nu} =0$ at $r=1$, we obtain a defect power law for$ \ell _\nu ^{\wedge}$:
\begin{equation}\label{eq:5.9}
\ell _\nu ^{\wedge} = \kappa (1 - {r^m})/m
\end{equation}
where $m=4$ for channel and TBL, and $m= 5$ for pipe, and $\kappa =1/(12c)$.

Note that \eqref{eq:5.8} is a diffusion equation; a heuristic derivation is presented below. For the generation of large scale eddies, the energy input is postulated $h({r})\ell _\nu ^{\wedge}({r})$ - in analogy to the external magnetic field $h(\bf {r})\Psi (\bf {r})$  as in \eqref{eq:5.1}; and energy output has two parts: one is the energy cascade to small scales postulated as $c{(\nabla \ell _\nu ^{\wedge})^2}$ - in analogy to $c{(\nabla \Psi )^2}$ in \eqref{eq:5.1}, and the other is the energy for spatial transport. As the bulk of turbulent flows is in statistical equilibrium (independent of time), the imbalance between $h\ell _\nu ^{\wedge}$ and $c{(\nabla \ell _\nu ^{\wedge})^2}$ determines the total turbulent transport, denoted as $G = \int {\Pi _p^ + d\bf {V}}  \propto \int {[h\ell _\nu ^{\wedge} - c{{(\nabla \ell _\nu ^{\wedge})}^2}]d\bf {V}} $. The minimum of $G$, through a variational calculation in analogy to \eqref{eq:5.2}, i.e. $0 = ({\bf{\tilde \delta }}G) \propto ({\bf{\tilde \delta }}\ell _\nu ^{\wedge})\int {[h + 2c\Delta \ell _\nu ^{\wedge}]d\bf{V}} $, where $\bf{\tilde \delta }$ denotes an infinitesimal variation, leads to \eqref{eq:5.8}.

The analytical expression for $\Theta$ is derived as follows. For TBL, $\Theta \approx 1$, as the quasi-balance extends all the way to the edge of the boundary layer (close to $\delta _{99} ^ +$), and the core layer is absent. For channel and pipe flow, $\Theta \approx c_1 /(r^2)$ in the center, because ${S^ + }{W^ + } \approx {c_2}{r^2} \to 0$ due to central symmetry, and a finite dissipation ${\varepsilon ^ + }\left( {r \to 0} \right) \approx {c_0}$ at centreline (given later herein), then a simple matching solution with no more parameter, is $ \Theta  = 1 - {c_1} + {c_1}{r^{ - 2}}$, valid through the quasi-balance region and the core layer with correct asymptotic conditions. This matching solution can also be derived by a simple parabolic distribution of turbulent transport, i.e.$\Pi _p^ +  = {S^ + }{W^ + }(\Theta  - 1) \approx {c_1}{c_2}(1 - {r^2})$. By denoting ${r_{core}} = \sqrt {{c_1}/(1 - {c_1})} $ which indicates the characteristic thickness for the core layer, we have:
\begin{equation}\label{eq:5.10}
{\Theta ^{CH\& Pipe}} = [1 + {({r_{core}}/r)^2}]/(1 + r_{core}^2)
\end{equation}
\begin{equation}\label{eq:5.11}
{\Theta ^{TBL}} = 1
\end{equation}

Therefore, the composite solution for the stress length function is obtained by substituting \eqref{eq:5.9}, \eqref{eq:5.10} and \eqref{eq:5.11} into \eqref{eq:5.7}. An interesting point is that the bulk flow constant $\kappa$ may be related to the Komolgorov constant $C_{k}$. This is because $\kappa =1/(12c)$, while the constant $c$ in \eqref{eq:5.8}, according to the above phenomenological picture, is postulated to indicate energy cascade into small scales. Therefore, $\kappa$ and $C_{k}$ may be closely related, since the energy cascade among different scales and the spatial momentum and energy transfer coexist.

Note that a further prediction is the turbulent dissipation in the outer flow, i.e. ${\varepsilon ^ + } = {S^ + }{W^ + }\Theta $. They are:
\begin{equation}\label{eq:5.12}
{\varepsilon ^{{ + {CH\& Pipe}}}} = \frac{{m{{({r^2} + r_{core}^2)}^{3/4}}}}{{\kappa {{{\mathop{\rm Re}\nolimits} }_\tau }(1 - {r^m}){{(1 + r_{core}^2)}^{3/4}}}}
\end{equation}
\begin{equation}\label{eq:5.13}
{\varepsilon ^{ + TBL}} = \frac{{4{{[1 - {\sigma ^{3/2}}{{(1 - r)}^{3/2}}]}^{3/2}}}}{{\kappa \sigma {{{\mathop{\rm Re}\nolimits} }_\tau }(1 - {r^4})}}
\end{equation}
Thus the centerline dissipation is ${\varepsilon _0} = m/[\kappa {{\mathop{\rm Re}\nolimits} _\tau }{(1 + 1/r_{core}^2)^{3/4}}$, which is ${\varepsilon _0} \approx 1.5/{{\mathop{\rm Re}\nolimits} _\tau }$ for pipe, and   ${\varepsilon _0} \approx 1.2/{{\mathop{\rm Re}\nolimits} _\tau }$for channel with $r_{core}=0.27$ at moderate \emph{Re}'s for DNS data. For large \emph{Re}'s, as the core layer thickness increases (see Part II and III), the centerline dissipation also increases, which is about ${\varepsilon _0} \approx 3.3/{{\mathop{\rm Re}\nolimits} _\tau }$  for pipe, and ${\varepsilon _0} = 2.6/{{\mathop{\rm Re}\nolimits} _\tau }$ for channel with $r_{core}=0.5$. These predictions await further verifications.

\section{Concluding Remarks}

In this paper, we present a derivation of a multi-layer formula for the stress length, motivated from an innovative Lie-group symmetry analysis. We find that the inner and outer MMEs admit dilation invariant solutions, expressed in terms of the stress length and its derivative. Three kinds of invariant solutions are proposed and validated by DNS data, in terms of three sets of parameters: scaling exponent, transition location (thickness in layered structure) and transition sharpness. These parameters are believed to be $Re$-independent at large $Re$'s (discussed in Part II and III).

A notable result from above analysis is the prediction of MVP, by substituting stress length function into the momentum balance equation. The solution for the mean shear for three canonical flows can be generally denoted as (hence MVP, friction coefficient etc can be obtained by integration)
\begin{equation}\label{eq:6.1}
S^ +  = ( - 1 + \sqrt {1 + 4\tau ^ + \ell _M ^ {+2}} )/2\ell _M ^{ +2}
\end{equation}
\begin{equation}\label{eq:6.2}
{U^ + } = \int_0^{{y^ + }} {{S^ + }d{y^ + }}
\end{equation}
where $\tau ^+$ is the total stress (for details see Part II and III). Thus, we have formulated a general framework: searching for group-invariant properties of relevant length order functions representative of the fluctuation structures (e.g. the stress length and some other length order functions), and then solving for the mean quantities (such as MVP) from balance equations.

What does the symmetry analysis using the Lie-group formalism add to our understanding beyond making a direct postulate of power law? The answer is that a Lie-group formalism guarantees that the RANS equation remains invariant under the (dilation) group of transformation, and any invariant solution ansatz used here guarantees that a transformed solution satisfies the RANS equation also. While the power law form for the stress length can be motivated from simple scaling arguments, two additional kinds would be difficult to imagine: a defect-power law form for the bulk flow and a series of matched forms derived from the generalized invariance ansatz. This last ansatz is determined by simple continuity and smoothness assumptions about the variation of the local group invariants; so they are the simplest possible more generalized invariant solutions.

The complexity of turbulent flows is that they generally encompass a hierarchy set of order functions, each displaying a specific type of turbulent transport. One will find other order functions responsible for turbulent transport of kinetic energy, temperature, etc. Ultimately, whether such closure solutions are true solutions need to be verified against empirical data. If verified, the length order function (and its gradient) is indeed the right quantity, and can be used as a valuable analysis tool. Note that the concept of the order function and the Lie-group formalism developed here are general and applicable to many other flow systems.

Note that identifying symmetries in order functions offer a way to `digest' the enriched experimental and DNS data. The present analysis opens a new direction for quantifying the mean profiles in a wide class of turbulent wall-bounded flows. Preliminary study has already applied it to compressible turbulent boundary layers \cite{zhangyousheng}, rough pipe \cite{shenjp}, pressure gradient effect, and buoyancy effect (Reyleigh-Benard convection) etc., to be communicated soon.

\section*{Acknowledgement}
We thank Brian Cantwell and Martin Oberlack for helpful discussions. Y. Wu has contributed during the initial stage of this work. This work is supported by National Nature Science Fund 11221062, 11452002 and by MOST 973 project 2009CB724100.

\appendix
\section{The structure ensemble dynamics (SED) theory }\label{appA}
\subsection{What is SED?}

The SED theory lays a conceptual foundation with the following system view of turbulence - both mean and fluctuations, being complementary to each other - governed by a single central element, viz., symmetry.

There are two kinds of complexities in turbulent flows that a unified theory of turbulence must incorporate: internal versus external complexities. One is related to the huge number of degrees of freedom, typically displayed by a continuous (power) spectrum over several decades of scales for velocity (or temperature, pressure) fluctuations. The second originates from the extreme sensitivity to external (boundary) conditions such that different geometry and different physical conditions (pressure gradient, roughness, buoyancy, rotation, etc) give rise to a large variety of flow structures (in practical flows). While the concept of energy cascade seems to form a universal basis for describing multi-scale phenomena (e.g. Kolmogorov (1941) theory and She-Leveque scaling \cite{she1994universal}), no concept has been proved viable to deal with the second kind of complexity. A consequence is that quantitative theories of technological flows are essentially empirical. The SED theory aims to discover a universal mechanism overarching the two complexities. It turns out that in wall turbulence, a well-defined mean profile always exists in the direction normal to the wall, which signifies the existence of an invariant distribution or statistical ensemble. The ensemble property is a quantifiable behavior of the system and embodies the simplicity behind the complexity, which, unfortunately, has never been investigated properly. For example, traditional approaches typically introduce average without specifying the ensemble. The motivation for using the word `ensemble' in the SED terminology makes this point explicit, and the wall turbulence is the first example where the wall introduces a preferred frame of reference defining the statistical ensemble.

Since symmetry is the central element governing both the mean and fluctuations, our system view then proposes a new way to solve the unclosed RANS equation. Instead of assuming an artificial relation between Reynolds stress (e.g. effect of fluctuations) and the mean shear with a specified form of the eddy viscosity in traditional approaches, we propose to identify the appropriate symmetry governing all terms in the RANS equations through an appropriate order function (e.g. stress length). In wall turbulence, a dilation symmetry is imposed by the presence of the wall (which breaks the Galilean invariance), and this symmetry is assumed to act on the stress length which is an interplay between the mean shear and Reynolds stress. This introduces a new dilation group of transformation which renders the RANS equation invariant. Note that the fact that the RANS equation is unclosed becomes irrelevant, since it is the solution manifold of the RANS equations under varying Re, Ma, pressure gradients, etc., which is leaved unchanged under the group of transformation. This last statement has not yet been mathematically proved, but the comparison to empirical data strongly supports its validation. Thus, a symmetry approach is the cornerstone of the SED theory.

The two fundamental properties of all physical systems are their space and time morphology; in turbulence, they are typically termed as structure and dynamics. Since turbulent fluctuations break all symmetries, the relevant symmetry in our discussion must belong to what is restored in the statistical sense \cite{Frischbook}. The SED theory affirms that universal symmetry exists in wall turbulence in the RANS equation, making Frisch's postulate of statistical symmetry restoration \cite{Frischbook} more explicit. On the other hand, the SED theory combines the universal Lie-group (symmetry) description with a multi-layer parameterization of spatial distributions originated from complicated structures and dynamics, which actual turbulent flows possess. The accomplishments reported in this series indicate that the SED is capable of resolving the theoretical challenge from the external complexity of actual flows. This is the origin of the three words in SED.

The three words (ensemble, structure, and dynamics), although abstract at this stage, lay down a theoretical framework for pursuing a viable theory of turbulence. SED aims to yield a quantitative theory of wall-bounded turbulent flows for spatial distributions of ensemble means of quantities such as velocity, Reynolds stress, kinetic energy, spectra, correlations, temperature, density, etc., and to identify relevant symmetries and physical constants giving rise to these quantitative means. In summary, the enduring postulate, validated through this series of work, is \textit{The time-averaged behavior of non-equilibrium systems such as wall-bounded turbulent flows generically possesses multiple statistical states, identifiable by order functions}. The two key concepts, i.e. multi-state and order function, are further explained below.
\par

\subsection{Multi-State symmetries with distinct energy balance mechanism}

Turbulence as a typical non-equilibrium process displays a number of symmetry-breaking. In wall-bounded flows, the variation along the direction normal to the wall forms the most outstanding spatial inhomogeneity. This variation is simple in a laminar state which comprises, for instance, a parabolic profile for a pipe and a Blasius profile in a laminar Boundary layer. In both cases, a single group of transformation suffices to describe the variation (see Section II for a Blasius profile). When turbulent fluctuations arise, several balance mechanisms in the energy dynamics (e.g. turbulence production or transport balances dissipation) are in competition, which is the origin of different scaling of the stress length. Thus, we highlight a key theoretical concept of multi-state as the most general form of the symmetry-breaking in wall turbulence.

A basic set of postulates are formulated as follows, to facilitate the generalization of the present analysis to other flows: (a) the existence of wall introduces a finite number of statistical states due to the presence of different characteristic fluctuation structures; (b) each state covers a spatially extended domain, which is a layer in wall-bounded flows, depending on the distance from the wall; and (c) layers, as well as transitions between layers, are characterized by the symmetry properties of the order functions. In other words, wall-bounded turbulent flows typically exhibit a `multi-layer structure'. Then, the key issue is to find appropriate variables and suitable formalisms for their descriptions; this is accomplished by the concept of order function as described below.
\par

\subsection{Order function}

Generally speaking, order functions are quantities displaying distinct symmetry property when significant inhomogeneity arises in space due to the presence of fluctuations. It is inspired by the concept of order parameter in the statistical mean-field theory, which describes the changes of the statistical state (e.g. phase transition) associated with symmetry-breaking \cite{kadanoff2009more}. In critical phenomena, the symmetry changes across a phase transition, which manifests in a change of the scaling exponent. In turbulent flow, fluctuations inherently alter the mean velocity (through Reynolds stress), and this interaction constitutes also a symmetry-breaking. This effect is described by introducing a length order function, which displays distinct character from one layer to another, by its dilation-invariant scaling. In a sense, turbulent fluctuations restore a dilation-symmetry (layer-by-layer) as theoreticians have guessed \cite{Frischbook}, and the symmetry property can be quantified by the scaling of the order functions, varying in space. The order functions also describe the variation of the flow to changes of the global parameters (such as $Re$, Mach number, Prandtl number, Rayleigh number, etc.) .

An order function typically involves a ratio of two (or more) statistical quantities. Finding an appropriate order function for a given turbulence system is the very first step in a SED study of turbulent flows. Three kinds of order functions have been suggested recently \citep{she2009}. The first is a ratio between two (dominant) terms in the governing equation (MME or MKE). This definition links the statistical state to its dynamical origin, in that the change of state is always attributable to the switching of the balance mechanism in the momentum or energy equation, easily displayed by ratio terms. The second is a length function, which is always a fundamental quantity for all physics phenomena; the complexity of turbulence lies in the fact that multiple characteristic length scales are relevant for different aspects of turbulence dynamics. Thus, appropriate length functions are particularly important for turbulence and dimensional argument is sufficient to define a number of relevant length scales (such as the stress length function). The third is a sensitive indicator function with correct (theoretical) asymptotic scaling that can be used to check the quality of simulation.

Note that we keep the specific definition of order function open, as additional fluctuations (such as density, temperature, etc.) may introduce new order functions. As more flows are studied, we will show that for each flow, there always exist a set of order functions which exhibit distinct symmetry behavior across different layers and hold important physical constants as Re increases. Some of the constants may even be universal for several kinds of flows. Then, the order function is a universal concept in the statistical analysis of inhomogeneous flows.
\par

\subsection{How does one proceed in a SED study of turbulence?}

A SED study of turbulence proceeds in three steps. First, it consists in verifying the existence of the symmetry, using empirical data. The theory asserts that for flows with well-defined mean profiles (e.g. stationary ensemble exists), there must exist a set of order functions representing the spatial symmetry. Then, analyzing the empirical behavior of the order functions (e.g. stress length) becomes a standard practice in the SED study. Symmetry test amounts to verifying that the order function has local scaling. Second, further analysis of empirical data determines the multi-layer parameters, such as scaling exponents and layer thicknesses. In particular, one would identify universal constants which do not vary with physical conditions (Re, Ma, geometry, pressure gradient, etc.). Finally, based on above qualitative (first step) and quantitative (second step) information, one evaluates important physical quantities (such as friction coefficient, heat flux, etc.) in the third step, for predictions. The three steps constitute a complete framework which not only identifies the basic symmetry of the flow, but also yields quantitative predictions of engineering interest. This procedure, and hence this framework, may be applicable to a number of flows with a variety of physical conditions (such as two phase flows, magneto-hydrodynamic flow, flows in tokamac, etc.).

The above procedure is particularly useful in renovating DNS study of practical flow systems, where analysis tools are scarce. Since the development of computation, voluminous quantity of data from DNS have been accumulated, but have not significantly improved our quantitative prediction power. Lack of an appropriate tool for extracting relevant information from data is the major bottleneck. The SED theory fulfills this need.

\section{ Basic concepts in the Lie group symmetry analysis}\label{appB}
\subsection {Lie group transformations}

A \textit{group} $G$  is a set of elements with a law of composition $\phi$ (an operator can be arbitrarily defined as addition, subtraction, multiplication etc., to be defined later) between elements satisfying the following axioms. For any element $a$, $b$ and $c$ of $G$, (1) Closure property: $\phi(a,b)$ is an element of $G$. (2) Associative property: $\phi(a,\phi(b,c))=\phi(\phi(a,b),c)$. (3) Identity element: there exists a unique element $i$ of $G$ such that $\phi(a,i)=\phi(i,a)=a$, and $i$ is called identity element. (4) Inverse element: there exists a unique element $a'$ in $G$  such that $\phi(a,a')=\phi(a',a)=i$ , and $a'$ is called inverse element.
\par
A \textit{transformation group } is defined as below. Let  $ \bf{x} = ({x_1},{x_2}, \ldots ,{x_n})$ lie in region  $D \in {R}^{n}$ (denotes n dimensional real space R). The set of transformations ${{\bf{x}}^*} = X({\bf{x}};\epsilon)$, depending on a parameter $\epsilon$ (not small) lying in set $S \in {R}$ , with  $\phi(\epsilon_1,\epsilon_2)$ defining a law of composition on the set $S$, forms a group of transformations if: (1) for each parameter $\epsilon$  the transformations are one-to-one onto $D$; (2) set $S$ follows the law of composition  $\phi$, forming a group $G$; (3)$\bf{{x}}^* = \bf{{x}}$  when $\epsilon =i$ , i.e. $X(\bf{{x}};i) = \bf{{x}}$; (4)$\bf{{x}}^*=X({\bf{x}};\epsilon)$  ,${{\bf{x}}^{**}} = X({{\bf{x}}^*};{\epsilon_2})$, then ${{\bf{x}}^{**}} = X({\bf{x}};\phi ({\epsilon_1},{\epsilon_2}))$. The last item is nontrivial and is justified on the basis that any twice transformations can be fulfilled by only one transformation. Imaging $\epsilon_1$ and $\epsilon_2$ are rotations, $\bf{{x}}^{**}$ reflect two successive rotations of $\epsilon_1$ and $\epsilon_{2}$ from initial $\bf{{x}}$.
\par
A \textit{one-parameter Lie group of transformation}  is defined in addition to satisfying the following axioms: (5) $\epsilon$ is a continuous parameter; (6) $X$ is infinitely differentiable with respect to $\bf{{x}}$ and an analytic function of $\epsilon$; (7)$\phi(\epsilon_1,\epsilon_2)$ is an analytic function of  $\epsilon_1$ and $\epsilon_2$. Here are some typical one-parameter Lie group transformations, namely translation, dilation, rotation and Galilean transformation, respectively:
\begin{equation}\label{eq:b1}
\left\{ \begin{array}{l}
x^*= x + \epsilon\\
y^*=y+ \epsilon
\end{array}\right .;\quad
\left\{ \begin{array}{l}
x^* = e^{\epsilon}x\\
y^* = e^{\epsilon}y
\end{array}\right .;\quad
\left\{ \begin{array}{l}
x^* = x\cos \epsilon  + y\sin \epsilon \\
y^* =  - x\sin \epsilon  + y\cos \epsilon
\end{array}\right .;\quad
\left\{ \begin{array}{l}
t^* = t\\
x^* = x + \epsilon t
\end{array}\right .
\end{equation}
Taking a two successive rotations as an example,
\begin{equation}\label{eq:b2}
\begin{array}{l}
\left\{ \begin{array}{l}
x^* = x\cos \epsilon_1  + y\sin \epsilon_1 \\
y^* =  - x\sin \epsilon_1  + y\cos \epsilon_1
\end{array} \right .  \\
\left\{ \begin{array}{l}
x^{**} = x^*\cos \epsilon_2 + y^*\sin \epsilon_2 = x\cos (\epsilon_1 + \epsilon_2) + y\sin (\epsilon_1 + \epsilon_2) \\
y^{**} =  - x^*\sin \epsilon_2 + y^*\cos \epsilon_2 =  - x\sin (\epsilon_1 + \epsilon_2) + y\cos (\epsilon_1 + \epsilon_2)
\end{array}\right.
\end{array}
\end{equation}
Therefore $\phi(\epsilon_1,\epsilon_2)=\epsilon_1+\epsilon_2$ , indicating any twice rotations can be fulfilled by only one rotation with parameter $\epsilon_1+\epsilon_2$ ; and $i=0$ (according to $\phi(a,i)=\phi(i,a)=a$ , that is  $a+i=a$ , therefore $i=0$), indicating no rotation. It can be easily verified that for all of the other transformations in \eqref{eq:b1}, the law of composition is  $\phi(\epsilon_1,\epsilon_2)=\epsilon_1+\epsilon_2$, and the identity is $i=0$
\par

\subsection{Group invariant and infinitesimal generator}

\textit{A group invariant (function, surface) } is what keeps unchanged under a transformation group, i.e. independent of parameter $\epsilon$ . Associated with each transformation, there is an invariant (any analytical function of such an invariant is also an invariant) of this transformation. When there is an invariant under transformation, we can say there is a symmetry. Specifically, if a function satisfies  $\Omega ({\bf{{x}}^*}) = \Omega (\bf{{x}})$ under translation, we say the function  $\Omega(\bf{{x}})$  is an invariant function admitting translation symmetry. Similarly, if any surface $\Omega(\bf{{x}})=0$  satisfies $\Omega(\bf{{x}}^{*})=0$ under translation, we say the surface is an invariant surface admitting translation symmetry. This is the same as for other transformations (dilation, rotation, Galilean transformation, etc.).

\textit{ Examples.} For the above translation, dilation, rotation and Galilean transformations, their group invariants - denoted as $\pmb{\rm{I}}(x,y)$  (a combination of variables by eliminating parameter $\epsilon$ ) are, respectively
\begin{equation}\label{eq:b3}
{\pmb{\rm{I}}} = y - x;\,\,{\pmb{\rm{I}}} = y/x;\,\,{\pmb{\rm{I}}} = x^2+y^2;\,\,{\pmb{\rm{I}}} =d^2 x/d t^2;
\end{equation}
One can check that all of them keep unchanged under corresponding transformations. Note that any analytical function $\Omega (\pmb{\rm{I}})$  is an invariant function, such as  $\pmb{\rm{I}}^2$ or $\sin (\pmb{\rm{I}})$ , etc.; and any surface  $\Omega(\pmb{\rm{I}})=0$ is also an invariant surface.

\textit{Infinitesimal transformations}. A general approach to obtain group invariant $\pmb{\rm{I}}(x,y)$  is by integrating the following characteristic equation:
\begin{equation}\label{eq:b4}
\frac{{dx}}{\xi } = \frac{{dy}}{\eta }
\end{equation}
where $\xi  = {\left. {{\partial _ \epsilon}{x^*}} \right|_{\epsilon = 0}}$  and $\eta  = {\left. {{\partial _ \epsilon}{y^*}} \right|_{\epsilon = 0}}$   are \textit{ infinitesimal transformations }for variable $x$  and $y$  respectively (see expansions on small parameter $\epsilon$ , i.e. ${x^*} = x + \epsilon \xi  + o({\epsilon^2})$  and  ${y^*} = y + \epsilon \eta  + o({\epsilon^2})$ ). The meaning of \eqref{eq:b4} is that infinitesimal increment of   $\pmb{\rm{I}}(x,y)$ is parallel (tangent) to the infinitesimal transformations $\xi$  and $\eta$ (in analogy for equations of a streamline $\pmb{\rm{I}}$ , where $\xi$  and $\eta$  are velocities in $x$  and $y$ , respectively), hence $\pmb{\rm{I}}(x,y)$  always keeps invariant under transformation. See later how these infinitesimal transformations are related to infinitesimal generators. The infinitesimal transformations for the above translation, dilation, rotation and Galilean transformations are:
\begin{equation}\label{eq:b5}
\left\{ \begin{array}{l}
\xi  = 1\\
\eta  = 1
\end{array} \right.;\quad
\left\{ \begin{array}{l}
\xi  = x\\
\eta  = y
\end{array} \right.;\quad
\left\{ \begin{array}{l}
\xi  = y\\
\eta  = -x
\end{array} \right.;\quad
\left\{ \begin{array}{l}
\xi  = 0\\
\eta  = t
\end{array} \right.
\end{equation}
Substituting them into the equation \eqref{eq:b4}, one obtains the group invariants in \eqref{eq:b3} by integration. Note that Galilean transformation is singular since $\xi =0$ , however, one can easily obtain the group invariant ${\pmb{\rm{I}}} = {d^2}x/d{t^2}$ , which means invariant of accelerated velocity.

\textit{Infinitesimal generator.} Above concepts can be extended to higher dimensional space, i.e. ${{\bf{x}}^*} = X({\bf{x}}{\bf{,y}};\epsilon)$ , ${{\bf{y}}^*} = Y({\bf{x}}{\bf{,y}};\epsilon)$, and to higher derivative space (so-called prolongation). Note that the infinitesimal transformations are uniquely determined by global transformations ($X$ and $Y$ ):
\begin{equation}\label{eq:b6}
\vec \xi ({\bf{x}},{\rm{y}}) = {\left. {{\partial _\epsilon}{{\bf{x}}^*}} \right|_{ \epsilon= 0}} = {\left. {{\partial _\epsilon}X({\bf{x}}{\rm{,y}};\epsilon)} \right|_{ \epsilon= 0}}
\end{equation}
\begin{equation}\label{eq:b7}
\vec \eta ({\bf{x}},{\rm{y}}) = {\left. {{\partial _\epsilon}{{\bf{y}}^*}} \right|_{ \epsilon= 0}} = {\left. {{\partial _\epsilon}Y({\bf{x}}{\rm{,y}};\epsilon)} \right|_{ \epsilon= 0}}
\end{equation}
in turn, they also determine the global transformations uniquely by integration:
\begin{equation}\label{eq:b8}
{\bf{x}}^* = e^{\epsilon \rm{\hat X}}{\bf{x}} = e^{\epsilon (\vec \xi \frac{\partial }{\partial \bf{x}} + \vec \eta \frac{\partial }{\partial \rm{y}})}\bf{x} = (1 + \epsilon \rm{\hat X} + \epsilon^2{\rm{\hat X}}^2/2! + \epsilon^3{{\rm{\hat X}}^3}/3! +  \ldots )\bf{x}
\end{equation}
\begin{equation}\label{eq:b9}
{\bf{y}}^* = e^{\epsilon \rm{\hat X}}\bf{y} = e^{\epsilon (\vec \xi \frac{\partial }{\partial \bf{x}} + \vec \eta \frac{\partial }{\partial \rm{y}})}\bf{y} = (1 + \epsilon \rm{\hat X} + \epsilon^2{\rm{\hat X}}^2/2! + \epsilon^3{\rm{\hat X}}^3/3! +  \ldots ){\bf{y}}
\end{equation}
where ${\rm{\hat X}} = \vec \xi \cdot {\partial _{\bf{x}}} + \vec \eta \cdot {\partial _{\rm{y}}}$ is the so called \textit{infinitesimal generator}. For example, by substituting \eqref{eq:b5} into the equations \eqref{eq:b8} and \eqref{eq:b9}, one can obtain the global transformations $X$ , $Y$ - explicitly given in \eqref{eq:b1},
Accordingly, the prolonged characteristic equations are
\begin{equation}\label{eq:b10}
\frac{{{\bf{dx}}}}{{\vec \xi ({\bf{x}},{\bf{y}})}} = \frac{{{\bf{dy}}}}{{\vec \eta ({\bf{x}},{\bf{y}})}} = \frac{{{\bf{d}}{{\bf{y}}^{[1]}}}}{{{{\vec \eta }^{[1]}}({\bf{x}},{\bf{y,}}{{\bf{y}}^{[1]}})}} =  \ldots  = \frac{{{\bf{d}}{{\bf{y}}^{[n]}}}}{{{{\vec \eta }^{[n]}}({\bf{x}},{\bf{y,}}{{\bf{y}}^{[n]}})}}
\end{equation}
where the super index [n] denotes the nth order derivative of dependent variable y, and${\vec \eta ^{[n]}}$  denotes infinitesimal transformations for $\bf{y}^{[n]}$ . All these equations can also be integrated to obtain Group invariants.
\par

\subsection{Symmetry transformation and invariant solution}

A \textit{symmetry transformation group} is a transformation group that keeps an ODE or PDE invariant. If a one-parameter Lie group of transformation ${{\bf{x}}^*} = X({\bf{x}}{\rm{,y}};\epsilon)$ , ${{\bf{y}}^*} = Y({\bf{x}}{\rm{,y}};\epsilon)$, leaves an equation ${C}(\bf{x},\bf{y})=0$ invariant, i.e. ${C}(\bf{x}^ *,\bf{y}^ *)=0$ , where  $\bf{x}$ and $\bf{y}$ are independent and dependent variables, respectively, then $X$ , $Y$ is a symmetry transformation group for the equation  ${C}(x,y)=0$ (or, we say that equation ${C}=0$  has a symmetry). For a given ODE or PDE, it is easy to verify whether some typical symmetry transformations as in \eqref{eq:b1} are permitted or not. In fact, there is a systematic way to calculate permitted symmetries. One can use mathematical software as \textit{Maple}, and also programs by \citet{Cantwellbook}, to carry out the calculations.

An \textit{invariant solution} is defined by the following two conditions: first, the solution satisfies the original ODE/PDE; second, it also keeps invariant under the symmetry transformation. Formally, let $K(\bf{x},\bf{y})=0 $ be an invariant solution of ${C}(\bf{x},\bf{y})=0$ corresponding to the symmetry transformation  ${{\bf{x}}^*} = X({\bf{x}}{\bf{,y}};\epsilon)$, ${{\bf{y}}^*} = Y({\bf{x}}{\bf{,y}};\epsilon)$, if and only if:

(i)  $K(\bf{x},\bf{y})=0 $ is a solution of the equation   ${C}(\bf x,\bf y)=0$.

(ii) The solution $K(\bf{x},\bf{y})=0 $ is invariant under transformation, i.e.$K(\bf{x}^ *,\bf{y}^ *)=0 $.
Note that (ii) means that,  $K(\bf{x},\bf{y})=0 $ must be rewritten as a function of group invariants, i.e.  $K(\pmb{\rm{I}}_1,\pmb{\rm{I}}_2,\ldots)=0 $. If there is additional variables $\bf{x}$  (or $\bf{y}$) in the expression of $K(\pmb{\rm{I}}_1,\pmb{\rm{I}}_2,\ldots)=0 $, then  $\bf{x}$  (or  $\bf{y}$ ) must change under transformation, thus breaking (ii). Therefore, one can use group invariants as variables to construct specific solutions. The so called candidate invariant solution by \cite{Oberlack2001} using constant group invariant assumption, automatically satisfies (ii) as a constraint; while for (i), its rationality lies in empirical verification using DNS and EXP data.


\bibliographystyle{jfm}

\begin{thebibliography}{99}

\bibitem{Smits2013}Smits, A. J. \& Marusic, I. 2013 Wall-bounded turbulence. Physics Today 66 (9), 25.

\bibitem{pope2000turbulent}Pope, S. B. 2000 Turbulent flows. Cambridge university press.

\bibitem{wilcox} Wilcox, D. C. 2006 Turbulence modeling for CFD. DCW industries La Canada, CA.
\bibitem{prandtl1925}Prandtl, L. 1925 Bericht uber die entstehung der turbulenz. Z. Angew. Math. Mech .

\bibitem{van1956} Driest, Van 1956 On turbulent flow near a wall. Journal of the Aeronautical Sciences
(Institute of the Aeronautical Sciences) 23 (11), 1007–1011.

\bibitem{karman1930} Von Karman, T. 1930 Mechanische ahnlichkeit und turbulenz, nachr. ges. wiss.
gottingen, math.-phys. kl.(1930) 58–76. Proc. 3. Int. Cong. Appl. Mech pp. 85–105.

\bibitem{marusic2010wall}Marusic, I., McKeon, B. J., Monkewitz, P. A., Nagib, H. M., Smits, A. J. \&
Sreenivasan, K. R. 2010 Wall-bounded turbulent flows at high reynolds numbers:
Recent advances and key issues. Physics of Fluids 22 (6), 065103.

\bibitem{millikan1938}Millikan, C. B. 1938 A critical discussion of turbulent flows in channels and circular
tubes. Proceedings 5th International Congress on Applied Mechanics .

\bibitem{Barenblatt}Barenblatt, G. I. 1993 Scaling laws for fully developed turbulent shear flows. Part 1. Basic hypotheses and analysis. {\it J. Fluid Mech.} {\bf 248}, 513-520.

\bibitem{Barenblatt19102004} Barenblatt, G. I. \& Chorin, A. J. 2004 A mathematical model for the scaling of turbulence. {\it Proceedings of the National Academy of Sciences of the United States of America.} {\bf 101}, 15023-15026.

\bibitem{Cipra17051996} {Cipra, B.} 1996 {A New Theory of Turbulence Causes a Stir Among Experts}. {\it Science}. {\bf 272}, {951}.

\bibitem{George05}George, W. K. 2005 Recent advancements toward the understanding of turbulent
boundary layers. in Proceedings of the Fourth AIAA Theoretical Fluid Mechanics Meeting,
Toronto, Canada (AIAA Paper No. 2005-4669).


\bibitem{Davidsonbook} {Davidson, P.A., Kaneda, Y., Moffatt, K. \& Sreenivasan, K.R.} 2011 {\it A voyage through turbulence.} Cambridge Univ. Press.


\bibitem{white} White, F. M. 2006 Viscous fluid flow. McGraw-Hill New York.

\bibitem{coles1956} Coles, D.
1956 The law of the wake in the turbulent boundary layer. {\it J. Fluid Mech.} {\bf 1}, 191--226.


\bibitem{Smits}Smits, A. J., McKeon, B. J. \& Marusic, I. 2011 High-reynolds number wall turbulence.
Annual Review of Fluid Mechanics 43 (1), 353–375.

\bibitem{Alfredsson2013} Alfredsson, P. H. and Imayamaa, S. and Lingwood, R. J. and Orlu, R. and Segalini, A. 2013 {Turbulent boundary layers over flat plates and rotating disks -The legacy of von Karman: A Stockholm perspective.} {\it European Journal of Mechanics B/Fluids}. {\bf 40}, 17-29.

\bibitem{monkewitz2007}Monkewitz, P. A., Chauhan, K. A. \& Nagib, H. M. 2007 Self-consistent highreynolds-
number asymptotics for zero-pressure-gradient turbulent boundary layers.
Physics of Fluids 19 (11).

\bibitem{nagib2008}Nagib, H. M. \& Chauhan, K. A. 2008 Variations of von krmn coefficient in canonical
flows. Physics of Fluids 20 (10).

\bibitem{nickels2004}Nickels, T. B. 2004 Inner scaling for wall-bounded flows subject to large pressure
gradients. Journal of Fluid Mechanics 521, 217–239.


\bibitem{javier2006} Del Alamo J.C. \& Jimenez J. 2006 Linear energy amplification in turbulent channels. {\it J. Fluid Mech.} {\bf 559}, 205--213.

\bibitem{Reynolds1972}Reynolds, W. C. \& Hussain, F. 1972 The mechanics of an organized wave in turbulent
shear flow. part 3. theoretical models and comparisons with experiments. Journal of
Fluid Mechanics 54, 263–288.

\bibitem{lvovprl} L'vov, V.S., Procaccia, I. \& Rudenco, O. 2008 Universal Model of Finite Reynolds Number Turbulent Flow in Channels and Pipes. {\it Phys. Rev. Let.} {\bf 100}, 050504(1-4).


\bibitem{Panton2007}Panton, Ronald L. 2007 Composite asymptotic expansions and scaling wall turbulence.
Philosophical transactions of The Royal Society A 365, 733–754.



\bibitem{Oberlack2001}Oberlack, M. 2001 A unified approach for symmetries in plane parallel turbulent shear
flows. Journal of Fluid Mechanics 427, 299–328.

\bibitem{lindgren2004} Lindgren, B., Osterlund, J.M. \& Johansson A.V. 2004 Evaluation of scaling laws derived from Lie group symmetry methods in zero-pressure-gradient turbulent boundary layers. {\it J. Fluid Mech.} {\bf 502}, 127-152.


\bibitem{marati2006mean}Marati, N., Davoudi, J., Casciola, C. M. \& Eckhardt, B. 2006 Mean profiles for
a passive scalar in wall-bounded flows from symmetry analysis. Journal of Turbulence
(7).
\bibitem{Oberlack2010new}Oberlack, M. \& Rosteck, A. 2010 New statistical symmetries of the multi-point
equations and its importance for turbulent scaling laws. Discrete Contin. Dyn. Syst.,
Ser. S 3, 451–471.

\bibitem{shenjp}She, Z. S., Wu, Y., Chen, X. \& Hussain, F. 2012 A multi-state description of
roughness effects in turbulent pipe flow. New Journal of Physics 14 (9), 093054.
\bibitem{bluman1991} Bluman, G. W. \& Kumei S. 1989 {\it Symmetries and differential equations} Springer-Verlag New York, Inc.

\bibitem{Cantwellbook} Cantwell, B.J. 2002 {\it Introduction to symmetry analysis.} Cambridge Univ. Press.


\bibitem{falkovich2009} Falkovich, G. 2009 Symmetries of the turbulent state. Journal of Physics A: Mathematical
and Theoretical 42 (12), 123001.

\bibitem{kadanoff2009more} Kadanoff, L. P. 2009 More is the same; phase transitions and mean field theories. {\it J. Stat. Phys.} {\bf 137}, 777-797.

\bibitem{Frischbook} Frisch, U. 1995 {\it Turbulence.} Cambridge Univ. Press.


\bibitem {cantwell1978a} Cantwell, B.J. 1978 Similarity transformations for the two-dimensional, unsteady, stream-function equation. {\it J. Fluid Mech.} {\bf 85}, 257-271.

\bibitem{cantwell1978b} Cantwell, B.J., Coles, D., \& Dimotakis, P.
1978 Structure and entrainment in the plane of symmetry of a turbulent spot. {\it J. Fluid Mech.} {\bf 87}, 641-672.



\bibitem{batchelor1951} Batchelor, G. K. 1951 Pressure fluctuations in isotropic turbulence.
Proc. Cambridge Philos. Soc. 47, 359.



\bibitem{wuyoups} Wu, Y., Chen, X., She, Z. S. \& Hussain, F. 2013 On the karman constant in turbulent channel flow. Physica Scripta 2013 (T155), 014009.


\bibitem{Iwamoto2002} Iwamoto, K., Suzuki, Y. \& Kasagi, N. 2002 Database of fully developed channel flow. {\it THTLAB Internal Report, No. ILR-0201}; see http://www.thtlab.t.utokyo. ac.jp.

\bibitem{hoyas2006} Hoyas, S. \& Jimenez, J. 2006 Scaling of the velocity fluctuations in turbulent channels up to $Re_\tau=2003$. {\it Phys. Fluids.} {\bf 18}, 011702; see http://torroja.dmt.upm.es/ftp/channels/.

\bibitem{wuxiaohua} Wu, X. H. \& Moin, P. 2008 A direct numerical simulation study on the mean velocity
characteristics in turbulent pipe flow. Journal of Fluid Mechanics 608, 81–112.

\bibitem{schlatter2010simulations}Schlatter, P., Li, Q., Brethouwer, G., Johansson, A. V. \& Henningson, D. S.
2010 Simulations of spatially evolving turbulent boundary layers up to reè= 4300.
International Journal of Heat and Fluid Flow 31 (3), 251–261.

\bibitem{wuyouscp} Wu, Y., Chen, X., She, Z. S. \& Hussain, F. 2012 Incorporating boundary constraints
to predict mean velocities in turbulent channel flow. SCIENCE CHINA Physics, Mechanics \& amp; Astronomy 55 (9), 1691.

\bibitem{schoppa2002coherent}Schoppa, W. \& Hussain, F. 2002 Coherent structure generation in near-wall turbulence.
Journal of fluid Mechanics 453, 57–108.


\bibitem{van1964} Van Dyke, Milton 1964 Perturbation methods in fluid mechanics, vol. 964. Academic
Press New York.

\bibitem{Landau1958} Landau, L.D. 1958 Statistical Physics. Pergamon Press.



\bibitem{she2009}She, Z. S., Chen, X., Wu, Y. \& Hussain, F. 2010 New perspective in statistical
modeling of wall-bounded turbulence. Acta Mechanica Sinica 26 (6), 847–861.

\bibitem{she1994universal}She, Z. S. \& Leveque, E. 1994 Universal scaling laws in fully developed turbulence.
Physical review letters 72 (3), 336.



\bibitem{zhangyousheng}Zhang, Y. S., Bi, W. T., Hussain, F., Li, X. L. \& She, Z. S. 2012 Mach-numberinvariant
mean-velocity profile of compressible turbulent boundary layers. Phys. Rev.
Lett. 109, 054502.

\setcounter{topnumber}{2} \setcounter{bottomnumber}{2}
\setcounter{totalnumber}{6}

\end{thebibliography}

$\\$ $^*$To whom correspondence should be addressed. E-mail:
she@pku.edu.cn

\end{document}